\PassOptionsToPackage{dvipsnames}{xcolor}
\documentclass[acmsmall,screen]{acmart}

\usepackage{tcolorbox}
\tcbuselibrary{breakable}
\usepackage{xspace}
\usepackage[inline]{enumitem}
\usepackage{caption}
\usepackage{subcaption}
\usepackage{bbding}
\usepackage{url}
\usepackage{booktabs}
\usepackage{tabularx}
\usepackage{circledsteps} 
\usepackage{graphicx}
\usepackage{footmisc}
\usepackage{soul}

\usepackage{listings}
\usepackage{newfloat}
\DeclareFloatingEnvironment[fileext=lol,name=Listing]{listing}
\lstdefinestyle{fancy}{
  basicstyle=\ttfamily\footnotesize,   
  keywordstyle=\color{blue},           
  commentstyle=\color{green!60!black}, 
  stringstyle=\color{red},             
  numbers=left,                        
  rulecolor=\color{black},             
  tabsize=2,                           
  breaklines=true,                     
  captionpos=b,                        
  showspaces=false,                    
  showstringspaces=false,              
  showtabs=false                       
}
\lstdefinelanguage{Markdown}{
  sensitive=false,
  keywordstyle=\color{blue},           
  commentstyle=\color{blue!70!black}\bfseries,
  morecomment=[l]{\#},
  moredelim=[s][\color{black!70!black}]{`}{`},
  literate=
    {*}{{{\color{black!70}*}}}1
    {-}{{{\color{black!70}-}}}1
}
\usepackage{multicol}

\usepackage{makecell}
\usepackage{multirow}
\newcolumntype{C}[1]{>{\centering\arraybackslash}p{#1}}

\usepackage{amsmath,bm,amsthm}
\theoremstyle{definition}

\usepackage{algorithm}
\usepackage{algpseudocode}
\algnewcommand\algorithmicforeach{\textbf{for each}}
\algdef{SE}[FUNCTION]{Function}{EndFunction}[2]{\algorithmicfunction\ \textnormal{#1}\ (#2)}{\algorithmicend\ \algorithmicfunction}%
\algdef{S}[FOR]{ForEach}[1]{\algorithmicforeach\ #1\ \algorithmicdo}
\algrenewcommand\alglinenumber[1]{\footnotesize #1}
\algrenewcommand\algorithmicrequire{\small \textbf{input:}}
\algrenewcommand\algorithmicensure{\small \textbf{output:}}
\algrenewcommand\algorithmicfunction{\textbf{Function}}
\algtext*{EndFunction} %
\algtext*{EndIf} %
\algtext*{EndFor} %
\algtext*{EndWhile} %

\newcommand{\LComment}[1]{\(\triangleright\) #1}
\AtBeginEnvironment{algorithmic}{\color{black}} 

\usepackage{fontawesome} 
\usepackage{framed}
\usepackage{textcomp}

\usepackage{microtype}
\usepackage{ifthen}

\title{\tool{}: Automated Metamorphic Test Generation via Functional Coupling Analysis}

\newcounter{rq}
\newenvironment{answertorq}{%
	\begin{tcolorbox}[
			arc=2mm,
			boxrule=0.5pt,
			left=2pt,
			right=2pt,
			top=2pt,
			bottom=2pt
		]
		\textbf{Answer to RQ\refstepcounter{rq}\therq{}:}
		}{\end{tcolorbox}}

\ifthenelse{\isundefined{\editmode}}{
	\newcommand{\editnote}[3]{%
	}

}
{
	\newcommand{\editnote}[3]{\xspace\colorbox{#1}{\sffamily \smaller \textcolor{white}{~\faCommenting{}~#2~}}\textcolor{#1}{~#3}\xspace}

}
\definecolor{latte-teal}{RGB}{23, 146, 153}
\definecolor{latte-sky}{HTML}{04A5E5}
\definecolor{nord0}{HTML}{2E3440}
\definecolor{nord1}{HTML}{3B4252}
\definecolor{nord2}{HTML}{434C5E}
\definecolor{nord3}{HTML}{4C566A}
\definecolor{nord4}{HTML}{D8DEE9}
\definecolor{nord5}{HTML}{E5E9F0}
\definecolor{nord6}{HTML}{ECEFF4}
\definecolor{nord7}{HTML}{8FBCBB}
\definecolor{nord8}{HTML}{88C0D0}
\definecolor{nord9}{HTML}{81A1C1}
\definecolor{nord10}{HTML}{5E81AC}
\definecolor{nord11}{HTML}{BF616A}
\definecolor{nord12}{HTML}{D08770}
\definecolor{nord13}{HTML}{EBCB8B}
\definecolor{nord14}{HTML}{A3BE8C}
\definecolor{nord15}{HTML}{B48EAD}

\newcommand{\code}[1]{\texttt{#1}}
\newcommand{\tool}{\textsc{MR-Coupler}\xspace}

\newcommand{\baseline}[1]{Baseline}

\newcommand{\deepSeek}{\textit{DeepSeek-V3.1}}
\newcommand{\deepSeekThink}{\textit{DeepSeek-V3.1-Think}}
\newcommand{\qwenThreeCoderFlash}{\textit{Qwen3-coder-Flash}}
\newcommand{\gptFourMini}{\textit{GPT-4o mini}}

\newcommand{\rqvalidity}{RQ1}
\newcommand{\rqbugrevealing}{RQ2}
\newcommand{\rqablation}{RQ3}
\newcommand{\rqsimilarity}{RQ4}
\newcommand{\rqcost}{RQ5}
\newcommand{\rqmrscout}{RQ6}

\newcommand{\relation}{\mathcal{R}}
\newcommand{\outputRelation}{\relation_o}
\newcommand{\inputRelation}{\relation_i}

\newcommand{\sourceInput}{x_{s}}
\newcommand{\followUpInput}{x_{f}}
\newcommand{\sourceOutput}{y_{s}}
\newcommand{\followUpOutput}{y_{f}}

\setcopyright{cc}
\setcctype{by}
\acmJournal{PACMSE}
\acmYear{2026} \acmVolume{3} \acmNumber{FSE} \acmArticle{FSE206}
\acmMonth{7} \acmDOI{10.1145/3808213}

\author{Congying Xu}
\orcid{0009-0000-2887-1690}
\email{congying.xu@connect.ust.hk}
\affiliation{%
  \institution{The Hong Kong University of Science and Technology}
  \city{Hong Kong}
  \country{China}
}
\affiliation{%
  \institution{Guangzhou HKUST Fok Ying Tung Research Institute}
  \city{Guangzhou}
  \country{China}
}

\author{Hengcheng Zhu}
\orcid{0000-0002-3082-5957}
\email{hzhuaq@connect.ust.hk}
\affiliation{%
  \institution{The Hong Kong University of Science and Technology}
  \city{Hong Kong}
  \country{China}
}
\affiliation{%
  \institution{Guangzhou HKUST Fok Ying Tung Research Institute}
  \city{Guangzhou}
  \country{China}
}

\author{Songqiang Chen}
\orcid{0000-0002-1220-8728}
\email{i9s.chen@connect.ust.hk}
\affiliation{%
  \institution{The Hong Kong University of Science and Technology}
  \city{Hong Kong}
  \country{China}
}
\affiliation{%
  \institution{Guangzhou HKUST Fok Ying Tung Research Institute}
  \city{Guangzhou}
  \country{China}
}

\author{Jiarong Wu}
\orcid{0000-0001-6126-303X}
\email{jwubf@connect.ust.hk}
\affiliation{%
  \institution{The Hong Kong University of Science and Technology}
  \city{Hong Kong}
  \country{China}
}
\affiliation{%
  \institution{Guangzhou HKUST Fok Ying Tung Research Institute}
  \city{Guangzhou}
  \country{China}
}

\author{Valerio Terragni}
\orcid{0000-0001-5885-9297}
\affiliation{%
  \institution{University of Auckland}
  \city{Auckland}
  \country{New Zealand}
}
\email{v.terragni@auckland.ac.nz}

\author{Shing-Chi Cheung}
\authornote{Shing-Chi Cheung is the corresponding author.}
\orcid{0000-0002-3508-7172}
\affiliation{%
  \institution{The Hong Kong University of Science and Technology}
  \city{Hong Kong}
  \country{China}
}
\affiliation{%
  \institution{Guangzhou HKUST Fok Ying Tung Research Institute}
  \city{Guangzhou}
  \country{China}
}
\email{scc@cse.ust.hk}

\begin{abstract}
    Metamorphic testing (MT) is a widely recognized technique for alleviating the oracle problem in software testing.
However, its adoption is hindered by the difficulty of constructing effective metamorphic relations (MRs), which often require domain-specific or hard-to-obtain knowledge.
In this work, we propose a novel approach that leverages the functional coupling between methods, which is readily available in source code, to automatically construct MRs and generate metamorphic test cases (MTCs).
Our technique, \tool{}, identifies functionally coupled method pairs, employs large language models to generate candidate MTCs, and validates them through test amplification and mutation analysis.
In particular, we leverage three functional coupling features to avoid expensive enumeration of possible method pairs, and a novel validation mechanism to reduce false alarms.
Our evaluation of \tool{} on 100 human-written MTCs and 50 real-world bugs shows that it generates valid MTCs for over 90\% of tasks, improves valid MTC generation by 64.90\%, and reduces false alarms by 36.56\% compared to baselines.
Furthermore, the MTCs generated by \tool{} detect 44\% of the real bugs.
Our results highlight the effectiveness of leveraging functional coupling for automated MR construction and the potential of \tool{} to facilitate the adoption of MT in practice.
We also released the tool and experimental data to support future research.

\end{abstract}

\begin{document}

\maketitle

\section{Introduction}

Metamorphic testing (MT) is a powerful testing technique that helps address the test oracle problem~\cite{segura2016survey, chen2018survey}.
Specifically, MT tackles this challenge by leveraging relations between multiple outputs (i.e., output relation) given the relations between corresponding inputs (i.e., input relation).
In this case, the logical implication from input relation to output relation defines a \emph{metamorphic relation (MR)}.
MT has been shown to be effective in various software systems, especially where the expected outputs are difficult to specify (e.g., large language models~\cite{cho2025llmorph,cho2025metamorphic}, compilers \cite{MT4Compiler_PLDI14SZD,MT4Compiler_OOPSLA16SZD}, machine translation \cite{jialunMTtranslator,csqMTtranslator}, question answering systems \cite{csqMTqa, shuaiMTqa}, etc.).
One MR can serve as an oracle applied to many test inputs to exercise diverse program behaviors and enhance test adequacy~\cite{mrscout}.

Despite its potential, the adoption of MT is challenging.
A key bottleneck is the construction of effective MRs~\cite{segura2016survey}, which requires domain-specific knowledge.
Although several attempts have been made to explore the generation of MRs, these approaches suffer from
\begin{enumerate*}[label*=(\roman*)]
\item reliance on manual effort~\cite{sun2016mumt, chen2016metric, DBLP:journals/infsof/ZhangCPTYZ25},
\item assumptions of regression testing scenarios~\cite{ayerdi2021generating,genmorph},
\item restriction to specific domains~\cite{zhang2014search, zhang2019automatic, DBLP:journals/infsof/ZhangCPTYZ25,DBLP:journals/tse/SeguraPTC18,zhou2018metamorphic,li2025mdpmorph,li2025mdpmorpi}, or
\item requirements for high-quality specifications~\cite{DBLP:conf/quatic/ShinPBB24,DBLP:journals/jss/BlasiGEPC21,mrscout}.
\end{enumerate*}
All these studies rely on knowledge that is hard to obtain.
Although a recent study \cite{mrscout} reported the possibility of mining the fragmented knowledge required by MRs from test cases, it also found such test cases to be rarely available:
they account for only 1\% of the studied test cases and are scattered in only 20\% of the studied projects.
The lack of automatic methodologies for constructing metamorphic test cases (MTCs) hinders the widespread adoption of MT.
To ease its adoption, a technique without the above-mentioned limitations is expected.
To construct such a technique, \emph{a central challenge is to formulate MRs without relying on knowledge that is hard to obtain}.

Fortunately, we found that the \emph{functional coupling between methods, which is readily available in the code, can be formulated as MRs}.
For example, the pair of functions \code{encrypt} and \code{decrypt} can formulate an MR $x=decrypt(encrypt(x))$, as shown in Listing \ref{lst:MTC-encryptDecrypt}.
This motivates us to formulate MRs by identifying such coupled method pairs.
This idea offers several advantages:
\begin{enumerate*}
\item \textit{readily-available knowledge}: it relies solely on a pair of methods and their implementation, which is by construction available in the scenario of unit testing;
\item \textit{more tractable problem}:
this transforms the challenging problem of deriving MRs into code understanding and relation reasoning, which can be effectively handled by current state-of-the-art large language models (LLMs)~\cite{mradopt,alphacode,arxiv23_FDUGPT4TestGen,TSE24_CompareGPTxSBST}.
For instance, although it is challenging to come up with MRs for a target method \code{encrypt},
when paired with a coupled method \code{decrypt}, it becomes easier for LLMs to understand their functionalities separately, realize that they are inverse functions, and then formulate a relation $ x=decrypt(encrypt(x))$;
\item \textit{easier bug manifestation}:
certain bugs can be revealed more easily with coupled computations.
For instance, while it is difficult to reveal the bug in Listing \ref{lst:code-encrypt-decrypt} by calling \code{encrypt} and \code{decrypt} separately, it becomes easier with the MR $x=decrypt(encrypt(x))$.
\end{enumerate*}
In summary, leveraging functionally coupled methods as a foundation for MR construction provides a practical and effective pathway to automate metamorphic testing and broaden its applicability.

However, leveraging functionally coupled methods to construct MRs requires addressing two technical issues.
First, given a target method, there can be dozens of candidate method pairs, and it is expensive to enumerate all possible method pairs blindly for MR construction.
Thus, there is a need for a precise mechanism to identify functionally coupled method pairs, which provides better focal methods for subsequent MR construction.
Second, while LLMs enable MTC generation via code understanding and reasoning, the resulting MTCs can be invalid due to hallucination~\cite{llmhallucination}.
Therefore, we need an effective mechanism to validate the generated MTCs, which allows us to avoid overwhelming developers with false alarms~\cite{gassert}.

To tackle these technical issues and effectively generate MTCs, we propose \tool, an automatic MTC generator for a given target method.
It operates in three phases.
First, it identifies functionally coupled methods as ingredients for MR construction, based on their signature commonality, functional call, and state interaction.
This design is informed by our observation of human-written MTCs and thereby addresses the first technical issue.
Second, it employs LLMs to generate MTCs based on each identified functionally coupled method pair by providing relevant and minimal context.
Specifically, we instruct LLMs to understand their functionalities and reason about potential MRs between them.
To reduce hallucinations that lead to invalid MTCs, we provide an MTC template and retrieve API usages for LLMs to follow.
Finally, it validates the generated candidate MTCs via test amplification and mutation analysis.
To validate the MTCs without a given ground truth, we create mutants from the original program by injecting artificial faults, and expect more amplified MTCs (from the candidate MTC) to pass on the original version compared with the faulty mutants.
This filtering strategy is based on a property of MT: the MR embedded in a correct MTC should apply to many other inputs to effectively kill mutants~\cite{mrscout}.

We evaluated \tool{} on (i) 100 human-written MTCs with corresponding target methods and (ii) 50 real-world bugs as tasks for evaluation.
\tool{} successfully generates valid MTCs for over 90\% of tasks, achieving a 64.90\% improvement in valid MTC generation and a 36.56\% reduction in false alarms compared with baselines.
The MTCs generated by \tool{} found 44\% of the 50 real bugs.
The key components play crucial roles in \tool{}: we found that functional coupling between methods guides LLMs to produce valid, bug-revealing MTCs, while MTC amplification and validation steps further improve bug detection and halve false alarms.
Last but not least, \tool{} can mimic developers in using functionally coupled methods and achieves over 90\% MR-skeleton consistency with human-written MTCs, highlighting its potential to assist developers in constructing MTCs.
In this paper, we make the following contributions.
\begin{itemize} [leftmargin=*]
	\item To the best of our knowledge, we are the first to construct MRs by leveraging the functional coupling between methods.
    Our approach relies solely on the code under test, and thus enables easier MR construction and lowers the barrier to adopting MT.
    \item  We propose \tool{}, an automatic approach to generate concrete metamorphic test cases.
    \tool{} instructs LLMs with relevant and minimal context for MR construction, and validates the generated MTCs with a novel filtering mechanism based on mutation analysis.
	\item We conduct extensive experiments to evaluate the effectiveness of \tool, including the validity of generated MTCs, the capability of revealing real bugs, and the similarity of generated MTCs to human-written MTCs.
	\item We make \tool{} and the experimental datasets publicly available.
    {The datasets of human-written MTCs and real-world bugs originally discovered by MT are valuable for future research.}
    Our artifact is available at the website of \tool{}~\cite{tool}.
\end{itemize}

\begin{listing}
	\centering
    \tiny
	\begin{lstlisting}[language=Java, basicstyle=\scriptsize\ttfamily]
@Test
public void testEncryptDecrypt() throws Exception {
    String plainText = "Hello AES!";
    SecretKey secKey = AESEncryption.getSecretEncryptionKey();
    //~Encrypt the plaintext: invoke on the source input, and produce the source output
    byte[] cipherText = AESEncryption.encryptText(plainText, secKey);
    // Decrypt the ciphertext: invoke on the follow-up input, and produce the follow-up output
    String decryptedText = AESEncryption.decryptText(cipherText, secKey);
    assertEquals(plainText, decryptedText); // output relation assertion
}
    \end{lstlisting}
	\vspace{-6pt}
	\caption{An MTC that Encodes the MR $x=decrypt(encrypt(x))$ over \code{encryptText} and \code{decryptText}}
	\label{lst:MTC-encryptDecrypt}
\end{listing}

\begin{listing}
	\centering
    \tiny
	\begin{lstlisting}[language=Java, basicstyle=\scriptsize\ttfamily]
public static byte[] encryptText(String plainText, SecretKey secKey) {
    Cipher aesCipher = Cipher.getInstance("AES");
    // aesCipher.init(Cipher.ENCRYPT_MODE, AESEncryption.defaultKey);   // bug
    aesCipher.init(Cipher.ENCRYPT_MODE, secKey);                        // fix
    return aesCipher.doFinal(plainText);
}
public static String decryptText(byte[] byteCipherText, SecretKey secKey) {
    Cipher aesCipher = Cipher.getInstance("AES");
    aesCipher.init(Cipher.DECRYPT_MODE, secKey);
    return aesCipher.doFinal(byteCipherText);
}
    \end{lstlisting}
    \vspace{-5pt}
    \caption{Code of Methods \code{encryptText} and \code{decryptText}~\cite{AESEncryptionTestGitHub}}
	\label{lst:code-encrypt-decrypt}
\end{listing}

\section{Preliminaries} \label{sec:preliminaries}

\label{sec:preliminary-MT}

Metamorphic Testing (MT) evaluates a program $P$ using a Metamorphic Relation (MR, $\relation$), which is a logical implication from an input relation $\inputRelation$ to an output relation $\outputRelation$~\cite{mrscout, segura2016survey, chen2018survey}.
\[ \relation: \relation_i \left( \sourceInput, \followUpInput \right) \implies \outputRelation \left( \sourceInput, \followUpInput, \sourceOutput, \followUpOutput \right) \]
$\inputRelation$ specifies the rule for deriving a \textit{follow-up input} ($\followUpInput$) from a given \textit{source input} ($\sourceInput$), while $\outputRelation$ defines the expected relationship over the inputs and their corresponding outputs ($\sourceInput$, $\followUpInput$, $\sourceOutput$, $\followUpOutput$)\footnote{{We incorporate both inputs and outputs in the output relation to present a more general definition of output relation that fits the MRs asserting the output relation based on inputs, such as round-trip translation~\cite{mrscout}.}}.

A \textit{Metamorphic Test Case (MTC)} is a test case that implements MT.
As illustrated in Listing \ref{lst:MTC-encryptDecrypt}, given an MR $\relation$ ($x=decrypt(encrypt(x))$) for a target program $P$ (class \code{AESEncryption}), an MTC has the following steps:
(i) constructing a source input $x_s$ (\code{plainText, secKey}),
(ii) executing $P$ (\code{AESEncryption.encryptText}) on $x_s$ to obtain the source output $y_s$ (\code{cipherText}),
(iii) constructing a follow-up input $x_f$ that satisfies $\relation_i$ (the source output $y_s$ also serves as the follow-up input),
(iv) executing $P$ (\code{AESEncryption.decryptText}) on $x_f$ to obtain the follow-up output $y_f$ (\code{decryptedText}),
and (v) verifying whether the output relation $\relation_o$ (\code{assertEquals(plainText, decryptedText)}) is satisfied.
Note that, following the definition by Xu et al.~\cite{mrscout}, the class \code{AESEncryption} is the \textit{target program} $P$, and \code{encryptText} and \code{decryptText} are the \textit{target methods} (the specific components of $P$ exercised during MT execution).
\code{encryptText} and \code{decryptText} are a pair of functionally coupled methods for formulating the MR $x=decrypt(encrypt(x))$.

MT offers several advantages:
\begin{enumerate*}[label*=(\roman*)]
\item \textit{Detecting faults in ``non-testable'' programs.}
MT is particularly valuable for validating \textit{non-testable programs} whose expected outputs for given inputs are difficult or infeasible to specify~\cite{segura2016survey, chen2018survey}.
Numerous empirical studies have demonstrated the effectiveness of MT in revealing faults in such scenarios~\cite{mrscout, mradopt, DBLP:conf/issta/MaS00C23, DBLP:journals/tse/SeguraPTC18, DBLP:conf/kbse/ChenJX21, DBLP:journals/tse/LiuKTC14}.

\item \textit{Reusable oracle across inputs.}
One MR can be applied to many inputs, enabling the test suite to exercise a broader range of program behaviors and thereby improve its fault-detection capability.
For example, in the AES case (Listing~\ref{lst:MTC-encryptDecrypt}), the MR $x = decrypt(encrypt(x))$ can be applied not only to an original input such as \code{"Hello AES!"}, but also to corner cases such as an empty string or strings containing special characters (e.g., \code{plainText="\textasciitilde{}!@\#\$\%\textasciicircum{}\&*()\_+"}).
\end{enumerate*}

\textbf{Goal.}
This motivates our work: developing a \textit{fully automated, domain-agnostic approach} to generate MTCs directly from code.
Our approach aims to bridge this gap by leveraging functional coupling between methods as the basis for formulating MRs, and by automatically generating valid MTCs that can be applied to diverse inputs to enhance test adequacy.

\section{Approach: \tool}

\begin{figure*}[t]
    \centering
    \includegraphics[width=.94\textwidth]{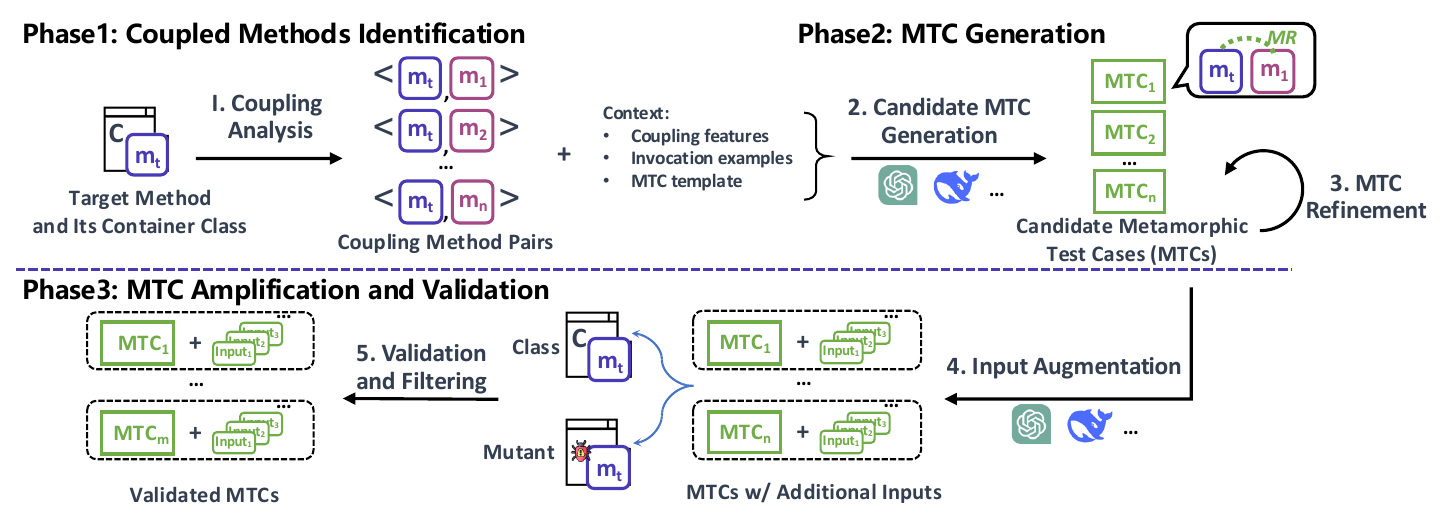}
    \vspace{-6.9pt}
    \caption{An overview of \tool} \label{fig:overview}
    \vspace{-5.5pt}
\end{figure*}

In this section, we present \tool{}, an automated MTC generator based on functional coupling.
Figure~\ref{fig:overview} presents an overview of \tool.
Given a target method and its container class as input, \tool produces a set of MTCs.
Specifically, the process consists of three phases:
\begin{enumerate}[leftmargin=*]
    \item \textit{Coupled Methods Identification}.
          In the first phase, \tool{} identifies functionally coupled methods that will be paired with the target method for MR construction.
          These methods are relevant to the target method in their intention or behavior.
          Such relevance can lead to potential MRs over their functionalities.

    \item \textit{MTC Generation}.
          In the second phase, \tool{} leverages LLMs to generate MTCs for the method pairs yielded by the first phase.
          To mitigate the impact of LLM hallucination \cite{llmhallucination}, we provide example usage of the involved methods and a template of MTC in the prompt.
          We also perform subsequent refinement based on the execution output of the generated MTCs.

    \item \textit{MTC Amplification and Validation}.
          The third phase amplifies the MTCs generated by the second phase with additional inputs.
          This phase serves two goals.
          First, this serves as a validation for the MTC based on a necessary property: a valid MTC with additional inputs (Listing~\ref{lst:LLM-newinputs-example}) should not have a lower pass rate on the original version than on the mutants.
          Second, the amplified MTCs can help reveal more bugs by exercising a broader range of program behaviors.
\end{enumerate}

\subsection{Phase 1: Coupled Methods Identification} \label{sec:app-relevance}

This phase is to identify the functionally coupled methods of the target method for MR construction.
However, given a target method and its container class, it is non-trivial to identify the functionally coupled methods that are suitable for MR construction.

On the one hand, classes often contain dozens of methods.
For example, in our dataset, there are over 30 candidate methods in a class for each target method on average.
Naively including all candidates not only increases the cost of LLM tokens but also risks of introducing noisy context, which distracts LLMs.
The difficulty is to identify those methods with relevance (e.g., producer–consumer, equivalent or inverse functions) suitable for MR construction.

\subsubsection{Characterization of Method Coupling.}

Although the number of possible methods is large, our investigation of human-written MTCs reveals that functionally coupled methods used for MR construction often exhibit coupling features in their \textit{Intent} (as reflected by method signatures) and/or \textit{Behavior} (as reflected by function calls and state interactions).

{
At the \textit{Intent} level, method couplings are often reflected by their signature commonality (i.e., common naming tokens, parameter types, and return types).
For example, the methods \code{encryptText} and \code{decryptText} in Listing~\ref{lst:code-encrypt-decrypt} have complementary intent: the former encrypts a plaintext string, while the latter decrypts a ciphertext.
This enables constructing an MR: $x=\mathit{decryptText}(\mathit{encryptText}(x))$.
Such coupling can be mechanically inferred from their signatures, and this also aligns with prior work on semantic coupling measurement via identifiers and comments~\cite{DBLP:journals/ese/PoshyvanykMFG09, DBLP:journals/infsof/FregnanBPB19}.}

{
Beyond the intent, coupled methods often exhibit \textit{Behavior} coupling through their function calls and state read/write interactions.
For example, the methods \code{insertElement} and \code{getElements} interact with the same object state via the field \code{element}: one updates it, while the other reads it.
An MR can be constructed as
$r \in \code{getElements}()~\textit{just after}~\code{insertElement}(r)$.
}
Similarly, behavioral coupling may also manifest in function calls, such as one method invoking the other or both calling the same functions.
{
Such a behavior-level coupling feature also aligns with prior work on structural coupling measurement via called functions and accessed variables~\cite{DBLP:journals/infsof/FregnanBPB19, DBLP:conf/scam/AllierVDS10}.
}
More examples of intent- and behavior-level coupling can be found in our artifact~\cite{tool}.

\subsubsection{Coupling Analysis.}
Motivated by the above coupling features observed from human-written MTCs,
in this step, \tool takes the target method $m_t$ and all the methods within its container class $\mathcal{C}$ as input, and identifies its functionally coupled methods. Each is combined with the target method to form a coupling method pair.
Algorithm~\ref{alg:coupling-analysis} summarizes the coupling analysis procedure.

\begin{algorithm}[t]
\footnotesize
\caption{Coupling Analysis for Identifying Functionally Coupled Methods}
\label{alg:coupling-analysis}
\begin{algorithmic}[1]
\Require Target method $m_t$, all methods $\mathcal{M}$ in the same class $\mathcal{C}$
\Ensure Coupled method pairs $\mathcal{M}_c$ and corresponding features $\mathcal{F}$

\State $\mathcal{M}_c \gets \emptyset$
\State $\mathcal{F} \gets$ empty mapping

\Statex \LComment{static features for each method}
\State $\mathsf{name}(m) \gets$ get the method name of $m$
\State $\mathsf{nameTok}(m) \gets$ tokenize method name of $m$
\State $\mathsf{paraRetType}(m) \gets$ get parameter and return types of $m$
\State $\mathsf{calls}(m) \gets$ get invoked APIs in $m$
\State $\mathsf{rField}(m),\mathsf{wField}(m) \gets$ get accessed/updated fields in $m$

\For{$m_i \in \mathcal{M} \setminus \{m_t\}$}
    \State $\mathcal{P} \gets \emptyset$
    \Statex \hspace{1.5em}\LComment{Feature1: Signature Commonality}
    \If{$\mathsf{name}(m_i) = \mathsf{name}(m_t)$}
        \State $\mathcal{P} \gets \mathcal{P} \cup \{\textsc{Intention}:\text{Overloading method}\}$
    \ElsIf{$\mathsf{nameTok}(m_i) \cap \mathsf{nameTok}(m_t)$ \textbf{and} $\mathsf{paraRetType}(m_i) \cap \mathsf{paraRetType}(m_t) \neq \emptyset$}
        \State $\mathcal{P} \gets \mathcal{P} \cup \{\textsc{Intention}:\text{Consuming/producing the same types of data type}\}$
    \EndIf
    \Statex \hspace{1.5em}\LComment{Feature2: Function Call}
    \If{$m_t \in \mathsf{calls}(m_i)$ \textbf{or} $m_i \in \mathsf{calls}(m_t)$}
        \State $\mathcal{P} \gets \mathcal{P} \cup \{\textsc{Behavior}:\text{Direct call dependency ($m_i\rightarrow m_t$/$m_t\rightarrow m_i$)}\}$
    \ElsIf{$\mathsf{calls}(m_i) \cap \mathsf{calls}(m_t) \neq \emptyset$}
        \State $\mathcal{P} \gets \mathcal{P} \cup \{\textsc{Behavior}:\text{Invoking the same APIs}\}$
    \EndIf
    \Statex \hspace{1.5em}\LComment{Feature3: State Interaction}
    \If{$\mathsf{wField}(m_i) \cap \mathsf{rField}(m_t) \neq \emptyset$ \textbf{or} $\mathsf{wField}(m_t) \cap \mathsf{rField}(m_i) \neq \emptyset$}
        \State $\mathcal{P} \gets \mathcal{P} \cup \{\textsc{State}:\text{Direct data dependency ($m_i\rightarrow m_t$/$m_t\rightarrow m_i$)}\}$
    \ElsIf{$\mathsf{rField}(m_i) \cap \mathsf{rField}(m_t) \neq \emptyset$ \textbf{or} $\mathsf{wField}(m_i) \cap \mathsf{wField}(m_t) \neq \emptyset$}
        \State $\mathcal{P} \gets \mathcal{P} \cup \{\textsc{State}:\text{Sharing the same dependent/dependency}\}$
    \EndIf

    \If{$\mathcal{P} \neq \emptyset$}
        \State $\mathcal{M}_c \gets \mathcal{M}_c \cup \{\langle m_t, m_i \rangle\}$
        \State $\mathcal{F}[m_i] \gets \mathcal{P}$
    \EndIf
\EndFor

\State \Return $(\mathcal{M}_c, \mathcal{F})$
\end{algorithmic}

\end{algorithm}
\vspace{8pt}

\textit{Feature1: Signature Commonality.}
To capture \emph{intent-level coupling}, \tool{} analyzes method signatures, including method names, parameter types, and return types (Lines~3--5 in Algorithm~\ref{alg:coupling-analysis}).
Given a target method $m_t$, a candidate method $m_i$ is considered if:
\begin{enumerate*}[label*=(\roman*)]
    \item $m_t$ and $m_i$ share the same method name; or
    \item $m_t$ and $m_i$ share common name tokens and overlap in parameter or return types (Lines~10--13).
\end{enumerate*}
Case (i) indicates method overloading, where an equivalence MR can be constructed.
{
For example, for the overloaded methods \code{base642bytes(String str)} and \code{base642bytes(String str, String code)}, an equivalence MR can be constructed as
$\code{base642bytes}(str)=\code{base642bytes}(str,code)$ (where $code$ is the default charset) ---
\textit{IF} the input string $str$ is converted to bytes without explicitly specifying a charset $code$,
\textit{THEN} the resulting byte array should be identical to the one obtained by explicitly using the default charset $code$~\cite{dubboGitHub}.
}
Case (ii) indicates the high likelihood that the two methods perform relevant functionalities.
{
For instance, for methods \code{encryptText(String plainText)} and \code{decryptText(byte[] byteCipherText, SecretKey secKey)}, an inversion MR can be constructed as
$x=\mathit{decryptText}$ $(\mathit{encryptText}(x))$.}
The matched features (e.g., shared naming tokens or parameters) are also provided to the LLM as contextual reference for the following MTC construction.

\textit{Feature2: Function Call.}
As method couplings may not always be reflected in signatures, \tool{} further captures \emph{behavior-level coupling} by analyzing function calls.
Specifically, it extracts the set of invoked APIs for each method (Line~6 in Algorithm~\ref{alg:coupling-analysis}) and checks:
\begin{enumerate*}[label*=(\roman*)]
    \item whether $m_t$ directly invokes $m_i$ (or vice versa), indicating call dependencies; or
    \item whether $m_t$ and $m_i$ share invoked APIs (Lines~14--17).
\end{enumerate*}
When $m_t$ directly invokes $m_i$ (or vice versa) and shows the call dependency between them, it may indicate a functionality specialization or composition relationship.
Besides, sharing common invoked APIs can indicate the commonality of their functionalities.
{
For example, \code{encryptText(String plainText)} can be viewed as a specialization of \code{encryptTextWithAbecedarium(String plainText, String abecedarium)}, as the former implicitly uses a default \code{abecedarium}.
Both methods invoke the same APIs (e.g., \code{Cipher.getInstance} in Listings~\ref{lst:code-encrypt-decrypt} and~\ref{lst:example of buggy AES}).
Accordingly, an equivalence MR can be constructed as
$\code{encryptText}(x)=\code{encryptTextWithAbecedarium}(x,a_{def})$ (where $a_{def}$ is the default \code{abecedarium}).}

\textit{Feature3: State Interaction.}
Beyond function-call behaviors, functional coupling may also be reflected by the state write/read interactions.
Therefore, \tool{} analyzes the fields read and written by each method (Line~7 in Algorithm~\ref{alg:coupling-analysis}).
Given $m_t$, a candidate method $m_i$ is considered if:
\begin{enumerate*}[label*=(\roman*)]
    \item $m_i$ writes fields that $m_t$ later reads (or vice versa), or
    \item $m_t$ and $m_i$ read or write the same fields, indicating shared-state access/update.
\end{enumerate*}

When $m_i$ updates fields that $m_t$ will read (or vice versa), this indicates a state interference between them.
Similarly, when they access or update the same fields, this can indicate the commonality of their data interaction. MRs can be constructed over such commonality.
{
For example, for two methods \code{insertElement} and \code{getElements}, they interact with the same field \code{element} of an object: the former updates it, whereas the latter reads it. Based on this, a contamination MR: $r\in\code{getElements}() just\ after$ $\code{insertElement}(r)$ can be constructed.
}

Given a target method, \tool{} analyzes all the methods within its container class, and identifies the coupled methods that match any of the features.
Note that a coupled method can match multiple features. The matching results are collected and then provided to LLMs as contextual reference for the following MTC construction (Listing~\ref{lst:LLM-mtc-generation}).

\subsection{Phase 2: MTC Generation} \label{sec:app-mtc-generation}

Given the set of coupling method pairs, this phase employs LLMs to come up with MRs and generate concrete MTCs that conduct MT.
However, generating valid MTCs remains challenging even with state-of-the-art LLMs~\cite{DBLP:conf/compsac/ZhangTP23, DBLP:conf/saner/ZhangSLD25}.
Many target methods require complex object instantiations, parameter configurations, or specific environmental setups, making valid method invocation difficult.
Without concrete method usage examples, LLMs are prone to hallucinations, e.g., producing code with non-existent APIs.
Moreover, generated MTCs must conform~to~the~steps~of~MT.

To address these challenges, \tool{} firstly provides LLMs with contextual guidance (e.g., method invocation examples and MTC template).
After prompting the LLMs to generate MTCs, \tool{} further refines them based on the execution output.

\begin{listing}[t]
    \centering
    \scriptsize
    \begin{lstlisting}[language=Markdown, basicstyle=\scriptsize\ttfamily]
You are an expert in Java programming and metamorphic testing, your task is to: ...
# Code of the paired method
    ```java
        byte[] encryptText(String plainText, SecretKey secKey) { ...
        String decryptText(byte[] byteCipherText, SecretKey secKey) { ...
    ```
# Coupling features on the paired methods
### Intention:
    * `encryptText` and `decryptText` operate on the same set of parameters and return types,
       but with different transformations.
        * `encryptText`: (String, SecretKey) -> byte[] , `decryptText`: (byte[],SecretKey) -> String
### Behavior:
    * both `encryptText` and `decryptText` invoke same APIs: `Cipher.getInstance(''AES'')` ...
# Invocation examples ...
# Skeleton of the container class ...
# Deliverable
    ```java
        public class $testClass${ ... <MTC Template> ... }
    ```
\end{lstlisting}
    \vspace{-5pt}
    \caption{Prompt Template for MTC Generation} \label{lst:LLM-mtc-generation}
\end{listing}

\subsubsection{Candidate MTC Generation.}
To help LLMs construct valid method invocations, \tool retrieves method invocation examples from the project under test and uses them as the guidance.
Specifically, for each method pair $\langle m_t, m_i \rangle$,
\tool searches for invocations of any of the two methods in the test code.
\tool scans the test files (under the \code{/test/} directory) in the project under test and uses \code{JavaParser}~\cite{urlJavaParser} to identify test methods annotated with \code{@Test}.
Each test is then checked whether it invokes $m_t$ or $m_i$; if so, it is collected as an invocation example.
{
To avoid excessive consumption of the LLM context window, \tool{} retains only the first three examples for each method and then provides them to the LLM as contextual guidance,} thereby increasing the likelihood that the generated tests correctly instantiate objects and invoke methods.

\textit{Prompt Design.}
Listing~\ref{lst:LLM-mtc-generation} shows a simplified prompt template for MTC generation.
Referring to the prompt design in recent studies~\cite{mradopt,arxiv23_FDUGPT4TestGen,DBLP:journals/corr/abs-2404-14646}, the prompt includes:
\begin{enumerate*}[label*=(\roman*)]
    \item a system message specifying the role of LLM and its tasks,
    \item the code of the paired methods,
    \item the coupling features of the paired methods,
    \item invocation examples,
    \item the skeleton of the container class,
    \item an MTC template that specifies the required deliverable.
\end{enumerate*}
This structured prompt provides both contextual information (ii–v) and task description (i and vi), guiding the LLM to generate syntactically correct MTCs.
The details of employed LLMs and their configuration can be found in Section~\ref{sec:evl-LLM-configuration}.
The output of this step is a set of candidate MTCs, each implemented as a standalone test class.

\begin{listing}[t]
    \centering
    \scriptsize
    \begin{lstlisting}[language=Java, basicstyle=\scriptsize\ttfamily]
public class AESEncryptionTest{
  @Test void MTC_input1() {
    String text = "Hello!"; var key = AESEncryption.getSecretEncryptionKey();
    byte[] encryptedText = AESEncryption.encryptText(text, key);
    String decryptedText = AESEncryption.decryptText(encryptedText, key);
    assertEquals(text, decryptedText);
  }
  @Test void MTC_input2() { String text = null; var key = null; ... }
  @Test void MTC_input3() { String text = "~!@"; var key = AESEncryption.getSecretEncryptionKey(); ... }
  @Test void MTC_input4() { String text = "_1234"; var key = AESEncryption.getSecretEncryptionKey(); ... }
  @Test void MTC_inputM() { String text = ""; var key = AESEncryption.defaultKey; ... }
}
\end{lstlisting}
    \vspace{-8pt}
    \caption{An amplified MTC (implementing an MR $x=\mathit{decryptText}(\mathit{encryptText}(x))$) with additional inputs} \label{lst:LLM-newinputs-example}
    \vspace{-8pt}
\end{listing}

\subsubsection{MTC Refinement.}\label{sec:app-refinement}
Consistent with prior observations~\cite{mradopt,DBLP:journals/corr/abs-2401-01701,ravi2025llmloop}, LLM-generated code frequently fails to execute, commonly due to errors such as \code{``cannot find symbol''}.
These errors typically arise from two sources:
\begin{enumerate*}[label*=(\roman*)]
    \item referencing non-existent classes, APIs, or fields, or
    \item missing dependencies (e.g., absent import statements).
\end{enumerate*}
To deal with this, \tool refines each non-compilable or non-executable MTC.
First, the error message is provided back to the LLM to request an automatically revised version.
If the revised MTC still fails to execute, \tool tries to fix the missing dependencies issue by statically analyzing the code using \code{JavaParser} to extract unresolved class names and searching for potential classes defined or imported in the project under test, and then adds the necessary import statements.
Finally, this phase results in a refined set of MTCs, which are subsequently used for amplification and validation.

\subsection{Phase 3: MTC Amplification and Validation} \label{sec:app-amplification-validation}
This phase validates candidate MTCs and diversifies their inputs.
Specifically, \tool{} amplifies each MTC with additional inputs and uses mutation analysis to refute invalid ones.
Specifically, mutants are created by injecting artificial faults, and a valid MTC with additional inputs (Listing~\ref{lst:LLM-newinputs-example}) is expected to pass more tests on the original program than on the mutated program.
In addition, the amplified MTCs can exercise a broader range of program behaviors, thereby increasing their bug-revealing capability.

\subsubsection{Input Augmentation} \label{sec:app-input-generation}

\tool{} amplifies each MTC by generating additional inputs to its MR to exercise a broader range of program behaviors, thereby enhancing its bug-revealing capability.
To generate these inputs, \tool employs LLMs by appending new instructions to the conversation for MTC generation and prompts the model to:
\begin{enumerate*}[label*=(\roman*)]
    \item review the previous conversation and context,
    \item review the previously generated MTC,
    \item apply its MR to new inputs by replacing the original input with $\langle$M$\rangle$ new inputs (such as boundary values, random data, or special characters) in the form of new test cases ($M=10$ by default), and
    \item output the new test cases within the same class following the naming convention (\code{MTC\_input1()}, \ldots, \code{MTC\_inputM()}).
\end{enumerate*}
Listing~\ref{lst:LLM-newinputs-example} shows a simplified example of an amplified MTC with 5 new inputs.

\subsubsection{Validation and Filtering} \label{sec:app-filtering}
Given an amplified MTC with additional inputs, \tool{} executes it on both the \emph{original} version of the target program and a \emph{mutated} version produced by injecting faults using \textsc{Major}~\cite{major}, and the pass rates $p$ (on the original version) and $p’$ (on the mutated version) are recorded.

A valid MTC with additional inputs is expected to pass more tests on the original program than on the mutated program.
When $p>p'$ or $p=p'=100\%$, the MTC is retained.
Considering a mutant (\code{encryptText} uses wrong \code{key} shown in Listing~\ref{lst:code-encrypt-decrypt}) and a valid MR $decrypt(encrypt(x))=x$ with additional inputs in Listing~\ref{lst:LLM-newinputs-example}, most inputs pass on the original version ($p=80\%$, except when \code{text=NULL} throws an illegal input exception) but fail on the mutant ($p'=20\%$, since only the case where \code{key = AESEncryption.defaultKey} coincidentally matches the default key succeeds). This MTC is retained.
By contrast, we observed an LLM-generated invalid MR $encrypt(plainText,secKey)=encryptTextWithAbecedarium(plainText,secKey, abecedarium)$.
When tested against a mutant where \code{encryptTextWithAbecedarium} fails to set up the \code{abecedarium} (Listing~\ref{lst:example of buggy AES}), most inputs in Listing~\ref{lst:LLM-newinputs-example} fail on the original version ($p=20\%$, except when the user-defined \code{abecedarium} happens to match the default).
On the mutant, all inputs pass ($p'=100\%$) because the \code{abecedarium} is ignored entirely.
Since $p<p'$, this MTC is filtered out.

When $p=p'=100\%$, two interpretations are possible:
\begin{enumerate*}[label*=(\roman*)]
    \item the injected mutants are ineffective and do not affect the tested behavior, or
    \item the MTC is ineffective in exposing mutants.
\end{enumerate*}
In such cases, \tool conservatively retains the MTC.

After validating each generated MTC, \tool finally outputs validated MTCs.

\begin{listing}
    \centering
    \scriptsize
    \begin{lstlisting}[language=Java]
public static byte[] encryptTextWithAbecedarium(
  String plainText,SecretKey secKey,String abecedarium) {
  Cipher aesCipher = Cipher.getInstance("AES");
  aesCipher.abecedarium = AESEncryption.abecedarium;         // bug
  // aesCipher.abecedarium = abecedarium;                    // fix
  aesCipher.init(Cipher.ENCRYPT_MODE, secKey);
  return aesCipher.doFinal(plainText);
}
    \end{lstlisting}
    \vspace{-10pt}
    \caption{Example of buggy \code{encryptTextWithAbecedarium}}
    \label{lst:example of buggy AES}
    \vspace{-8pt}
\end{listing}

\section{Evaluation}
In this section, we present our evaluation of \tool{}.
Specifically, we aim to answer the following research questions (RQs).

\begin{itemize}
	\item[\textbf{\rqvalidity{}}] \textbf{Validity:} \textit{How effective is \tool at generating MTCs?}
          This RQ investigates the overall effectiveness of \tool{}, in terms of the validity and correctness of the generated MTCs.
          In addition, we compared it with vanilla-LLM-based baselines to understand its superiority.
	\item[\textbf{{\rqbugrevealing{}}}] \textbf{Bug-Revealing Capability:} \textit{How effective is \tool{} in revealing real-world bugs?}
            Compared to seeded bugs (e.g., mutants), real-world bugs are often more sophisticated.
            Thus, we evaluate whether the MTCs generated by \tool{} can detect real-world bugs as the human-written MTCs do.
	\item[\textbf{\rqablation{}}] \textbf{Ablation Study:} \textit{How does each step contribute to the effectiveness of \tool?}
          \tool{} consists of three key steps: \emph{Coupling Analysis}, \emph{Input Augmentation}, and \emph{Validation and Filtering}.
	       This RQ performs an ablation study to assess the contribution of each step.

	\item[\textbf{\rqsimilarity{}}] \textbf{Similarity:} \textit{Do the MTCs generated by \tool{} share the same MR skeletons as human-written ones?}
    This RQ evaluates whether \tool{}-generated MTCs align with developers' practices in selecting coupled methods and constructing input/output relations.

	\item[\textbf{\rqcost{}}]
	{
	\textbf{Cost:} \textit{What is the cost of \tool{} in practice?}
	This RQ investigates the computational cost of \tool{} in terms of LLM token consumption and monetary expenditure. This is essential for assessing the practical scalability of \tool{} in the real world.
	}

	\item[\textbf{\rqmrscout{}}]
	{
	\textbf{Comparison Study:} \textit{How does \tool{} outperform MR-Scout?}
    MR-Scout~\cite{mrscout} is a recent technique to synthesize MRs from MTCs.
    To understand how \tool{} compares to MR-Scout, we contrasted the two along \textit{applicability}, \textit{scalability}, and \textit{correctness}.
    }

\end{itemize}

\subsection{Datasets} \label{sec:dataset}
We prepared two datasets to answer the above RQs. The first dataset includes pairs of target methods together with corresponding human-written MTCs available in open-source projects to evaluate the validity and similarity of the generated MTCs (\rqvalidity{} and \rqsimilarity{}). The other is a subset from the first dataset, including only the cases whose MTCs can reveal bugs on a historical buggy version of the target program, for evaluating the bug-revealing capability of generated MTCs (\rqbugrevealing{}).

\subsubsection{Human-Written MTCs}
The first dataset contains 1,471 MTCs written by developers in open-source Java projects.
Each entry in this dataset consists of a human-written MTC and a corresponding pair of MR-coupled methods.
These MTCs are valid and executable.
Such a dataset is leveraged to (i) evaluate the validity of automatically generated MTCs (\rqvalidity{}), by running them on executable target methods, and (ii) measure the similarity between the human-written MTCs and \tool generated MTCs, by checking whether they encode the same MR-skeletons (\rqsimilarity{}).

To construct such a dataset, we adopted a strategy similar to Xu et al.~\cite{mrscout}.
Specifically, we collected a list of high-quality Java projects (i.e., with at least 50 stars) from GitHub.
The query was done on {December-16, 2024}, which returned over {24,000} projects.
We then ran \textsc{MR-Scout}~\cite{mrscout} to discover human-written MTCs from these projects, which yielded {46,006} candidate MTCs.
{With these candidates}, we applied three filtering criteria to select the valid and executable MTCs:
\begin{enumerate*}[label=(\roman*)]
\item they must compile, {as we need to compile and run the test cases in our experiments};
\item they must pass in the latest version of the project {to ensure the MTCs are valid}; and
\item {the commit introducing these MTCs must mention an issue number in its commit message} (e.g., containing ``\code{\#123}''). We prioritize such tests since they are often extensively discussed and reviewed to disclose the issues and thus tend to be of high quality.
\end{enumerate*}

The whole process yielded a dataset containing 1,471 entries.
Each entry in this dataset consists of a human-written MTC and a corresponding pair of MR-coupled methods.
{
The 1,471 MTCs come from 213 Java projects, which on average contain 3,505.44 commits, 1,772.73 Java files, and 199,602.38 lines of Java code.
Detailed information about these MTCs and projects is available on the \tool{} website~\cite{tool}.
}
We ran \textsc{MR-Scout}~\cite{mrscout} to obtain the corresponding pair of MR-coupled methods for each MTC.
For example, in the MTC that encodes the relation $x = decrypt(encrypt(x))$ (Listing~\ref{lst:MTC-encryptDecrypt}), \code{encryptText} and \code{decryptText} are the MR-coupled methods and will be identified by \textsc{MR-Scout}.
We use the first invoked method \code{encryptText} as the target method and take \code{decryptText} as the ground truth of a coupled method.
{Each entry formulates an MTC generation task used for our experiments of \rqvalidity{} and \rqsimilarity{}}.

\subsubsection{Bug-revealing MTCs.} \label{sec:dataset-Bug-revealing MTCs}
The other dataset is made up of 50 entities with bug-revealing MTCs filtered from the first dataset. These entities are used to evaluate whether \tool{} can generate effective MTCs to reveal real bugs as the human-written MTCs do (\rqbugrevealing{}). We identify such entities from the first dataset by checking whether their MTC fails on a buggy version and passes on a fixed version. Specifically, an issue report may be resolved through commits; thus, for each issue-associated MTC, we identify two versions of the project:
\begin{enumerate*}[label=(\roman*)]
\item the potential \emph{buggy version}, defined as the commit before all issue-related commits;
\item the potential \emph{fixed version}, defined as the last commit of all issue-related commits.
\end{enumerate*}
We then execute the MTC on both versions.
We consider an MTC bug-revealing only if it fails on the buggy version and passes on the fixed version.

This process required significant manual effort to set up specific project environments for multiple versions of each project and resolve complicated dependency issues.
We ultimately reproduced 50 MTC-bug pairs, obtaining their corresponding buggy and fixed program versions, which form the benchmark for evaluating the bug-revealing capability of \tool{} (\rqbugrevealing{}).

\subsection{{Evaluation Setup}}\label{sec:evl-setup}
This section introduces the employed LLMs, baselines, and the experiment environment.

\subsubsection{Employed Large Language Models}
\label{sec:evl-LLM-configuration}

\tool employs LLMs to generate MTCs and their {alternative inputs}.
In the evaluation, we include representative state-of-the-art LLMs~\cite{urlEvalLeaderboard}, covering general-purpose, coding, and reasoning LLMs from well-known model families.
Specifically, they are {\gptFourMini} from OpenAI~\cite{urlGPTFourMini},{\qwenThreeCoderFlash} from Alibaba~\cite{urlCodeQwen}, {\deepSeek} and {\deepSeekThink} from DeepSeek~\cite{urlDeepseek}.
Following a typical setup in recent studies~\cite{mradopt,DBLP:conf/icse/Du0WWL0FS0L24,DBLP:journals/corr/abs-2403-16898}, for each MTC generation task, we repeated the generation process five times with a temperature setting of 0.2.

\subsubsection{Baselines}
To the best of our knowledge, there is no existing fully automated and domain-agnostic approach to generate metamorphic test cases for a given program under test.
{Although some approaches are proposed to generate domain-specific MRs~\cite{DBLP:journals/infsof/ZhangCPTYZ25,zhang2019automatic}, or synthesize MRs based on human-prepared materials~\cite{mradopt,mrscout,nolasco2024} or manual effort~\cite{DBLP:conf/quatic/ShinPBB24} (discussed in Section~\ref{sec:relatedwork}), adapting them into comparable automated domain-agnostic baselines is non-trivial.}
Given the proven effectiveness of LLMs in code~\cite{mradopt,alphacode,ChainOfCode} and test generation~\cite{TSE24_CompareGPTxSBST, arxiv23_FDUGPT4TestGen}, we set \textit{directly prompting LLMs} as a baseline.
In this baseline, we allow LLMs to conduct a round of revision to the generated code based on the execution feedback as in our method, which is found to be an effective common post-processing to enhance code generation~\cite{arxiv23_FDUGPT4TestGen}.
The baseline uses a similar prompt template (Listing~\ref{lst:LLM-mtc-generation}), and follows the same refinement step in Section~\ref{sec:app-refinement}.
Experimental results show that a target method can be paired with 6.93 relevant methods (rounded to 7) by \tool{} on average.
Therefore, to have a fair comparison, we instructed the baseline LLMs to generate 8 candidate MTCs, which correspond to the target method itself, plus the 7 additional relevant methods.
For each task, we repeat the generation process five times (temperature: 0.2), consistent with \tool's configuration.

\subsubsection{Experimental Environment.} All experiments were conducted on a machine with a 64-core AMD Ryzen Threadripper PRO 3995WX CPU and 512 GB RAM.
The LLMs in our evaluation are running on cloud platforms and accessed via the official APIs of OpenAI, Alibaba, and DeepSeek.

\subsection{\rqvalidity{}: Validity of Generated MTCs} \label{sec:rq-validity}

\rqvalidity{} aims to evaluate the overall effectiveness of \tool in generating valid MTCs.
To this end, we run \tool{} on the target methods in the dataset of human-written MTCs.
We also compare it against the baselines (Section~\ref{sec:evl-setup}).

\subsubsection{Experiment Setup}
Experiments were conducted at scale using four LLMs, with each MTC generation task repeated five times (Section~\ref{sec:evl-LLM-configuration}).
Running \tool{} and the baselines on all 1,471 tasks is time-consuming and unaffordable.
Therefore, we evaluated \tool{} and baselines on 100 randomly sampled entries.
This sample size ensures a 95\% confidence level with a margin of error \(\le 10\%\).
The randomly sampled 100 entries come from 55 projects. The details are available on \tool{}'s site~\cite{tool}.
For each entry (i.e., target method), \tool{} generates multiple MTCs. We present the total number of generated test cases, and measure the effectiveness using the following metrics:

\begin{itemize}[leftmargin=*]
	\item \textit{Percentage of Executable MTC:} The proportion of generated test cases that compile, run without error, and satisfy the two necessary properties of an MTC~\cite{mrscout}: (i) at least two invocations of target methods; (ii) one assertion relating those invocations' inputs and outputs.
    \citet{mrscout} reported, MR-Scout can achieve 97\% precision, and we ran MR-Scout on each case.
	\item \textit{Percentage of Valid MTC:} The proportion of valid MTCs to all generated test cases, where a valid MTC is an executable MTC that passes on the latest project version. We assume the latest version of a target method is of low probability to be buggy, as it has passed human-written MTCs.
	\item \textit{Number of Successful Tasks:} The number of target methods generated with at least~one~valid~MTC.
	\item \textit{Percentage of False Alarm:} The proportion of invalid MTCs to all executable MTCs, where an invalid MTC satisfies the properties but fails on the latest version.
\end{itemize}

\subsubsection{Experiment Results.}

\begin{table}[t]
	\scriptsize
	\caption{Effectiveness of \tool in Generating Valid MTCs for 100 Target Methods}\label{tab:rq-validity}
	\centering
    \vspace{-8pt}
	
\setlength{\tabcolsep}{2.5pt}
\begin{tabular}{l|rr|rr|rr|rr|r}
	\toprule
	\multirow{2}{*}{Metric} & \multicolumn{2}{c|}{GPT-4o-mini} & \multicolumn{2}{c|}{Qwen3-coder-Flash} & \multicolumn{2}{c|}{Deepseek-V3.1} & \multicolumn{2}{c|}{Deepseek-V3.1-Think} & \multirow{2}{*}{\textbf{Improv.}}                                                        \\
	\cmidrule{2-3}\cmidrule{4-5}\cmidrule{6-7}\cmidrule{8-9}
	                        & Baseline                         & \tool                                  & Baseline                           & \tool                                    & Baseline                          & \tool     & Baseline  & \tool                        \\
	\midrule
	Num. of Generated TCs           & 3923                             & 4176                                   & 3984                               & 3911                                     & 3968                              & 3626      & 3971      & 4151      & {-}              \\
	Pct. of Executable MTC     & {50.40\%}                        & {83.69\%}                              & {60.02\%}                          & {92.66\%}                                & {60.26\%}                         & {93.08\%} & {62.08\%} & {88.44\%} & \textbf{54.35\%} \\

	Pct. of Valid MTC          & {40.66\%}                        & {71.84\%}                              & {47.52\%}                          & {80.98\%}                                & {52.60\%}                         & {84.94\%} & {55.35\%} & {83.57\%} & \textbf{64.90\%} \\
	Num. of Successful Tasks      & {62}                           & {92}                                 & {76}                             & {91}                                   & {82}                            & {95}    & {85}    & {98}    & \textbf{24.82\%} \\

	\midrule
	Pct. of False Alarm        & 19.32\%                          & 14.16\%                                & 20.70\%                            & 12.61\%                                  & 12.71\%                           & 8.74\%    & 10.83\%   & 5.50\%    & \textbf{36.56\%} \\

	\bottomrule
\end{tabular}

\end{table}

Table~\ref{tab:rq-validity} shows the results of \tool in generating valid MTCs for 100 target methods.
\tool generated valid MTCs for over 90 target methods, achieving its best performance with \deepSeekThink{}. \tool{} produced valid MTCs for 98 of 100 target methods with a 5.5\% false alarm rate.
Compared to baseline, \tool{} improves the valid MTC rate by 64.90\% and reduces false alarms by 36.56\%.
Even with the weakest model (\gptFourMini), \tool{} outperforms baselines using stronger models (e.g., \deepSeekThink).

The improvements are from two key factors.
On the one hand, providing LLMs with functionally coupled methods serves as a hint, effectively inspiring them to infer valid MRs and then generate valid MTCs, thereby reducing hallucinations.
Without such context, coming up with MRs from scratch is challenging for LLMs, leading to higher rates of invalid MRs.
On average, \tool identified 6.93 relevant methods per target method from 30.44 candidate methods in their container classes, significantly narrowing the enumeration space.
On the other hand, retrieving real invocation examples helps LLMs construct valid input objects and correctly invoke methods, particularly when methods require complex object instantiations.
This context helps generate 83.69\% to 93.08\% executable MTCs, which are 54.35\% more compared with baselines.

\textit{Failure analysis.}
{
When built with \deepSeekThink{}, \tool fails to generate any valid MTCs for two target methods.
This is because, for one target method, execution requires access to external or environmental resources (e.g., a JSON file or environment variable),} but no invocation examples were available in the repository as the context.
As a result, \tool failed to configure these resources in the generated MTCs, leading to non-executable tests and no valid MTCs.
For the other target method, \tool generated all executable but invalid MTCs (i.e., false alarms).

\tool exhibits a 5.5\% false-alarm rate when built with \deepSeekThink{}. There are two main reasons.
\begin{enumerate*}[label*=(\roman*)]
    \item
    \tool relies on \textsc{Major}~\cite{major} to generate mutants for validation (Section~\ref{sec:app-filtering}).
    For some target methods, \textsc{Major} failed to execute due to environmental issues (e.g., uncompilable dependencies), preventing mutant generation and disabling the validation step that filters false alarms.
	\item
    For some cases, the generated MTCs cannot reveal the injected mutants, and do not expose behavioral differences between the base and mutated versions -- the pass rates are identical, causing \tool to retain false alarms.
\end{enumerate*}

\subsubsection{{Correctness of Generated MTCs}}\label{sec:evl-correctness}
While the above automatic and objective metrics evaluate the validity of generated MTCs by checking the necessary properties and their execution results, this does not guarantee the semantic correctness of the underlying MRs.
A generated MTC may satisfy these validity criteria yet still be semantically incorrect or meaningless (e.g., vacuous oracles).
To assess the semantic correctness, we further conduct a manual study to validate whether generated MTCs indeed reflect correct MRs for the target method.

{
\textit{Setup.}
For each of the 100 target methods in RQ1, we manually inspected one randomly selected \tool-generated MTC.
Two participants (each with over four years of experience in Java and metamorphic testing) independently validated each selected MTC. For cases with disagreement, the two participants discussed and reached a consensus.
}

An MTC is deemed \emph{correct} if it reflects an MR present in the code; it may miss failures but must not produce false alarms.
For each MTC, participants first understand the functionality of the target method and its coupled method.
They then analyzed the underlying MR an MTC and determined whether an MTC is
correct. Particularly, they assess whether any valid input could violate the assertion and checked execution for false alarms.
If any of the cases is found, the MTC is incorrect.

{
\textit{Result.}
Among the 100 sampled MTCs, 8 were \emph{Incorrect}, and 3 are too domain-specific and complicated to understand and validate their correctness.
The incorrect cases mainly involved faulty test implementations (e.g., triggering \code{IllegalArgumentException} or \code{NullPointerException}) or MRs not holding for corner-case inputs. The remaining 89 MTCs were \emph{Correct}.
The inter-rater agreement between participants was high (87\%), and disagreements were resolved through discussion.
Overall, these results suggest \tool{}'s effectiveness in generating correct MTCs.
}

\begin{answertorq}
    \tool successfully generates valid MTCs for over 90\% of tasks, achieving 64.90\% and 36.56\% improvements in generating valid MTCs and reducing false alarms, respectively, compared with baselines.
    {Our manual validation shows that only 8 out of 100 sampled generated MTCs were incorrect.}
\end{answertorq}

\subsection{\rqbugrevealing{}: Bug-revealing capability} \label{sec:rq-bug-revealing}
\begin{table}[t]
	\scriptsize
	\caption{Bug-Revealing Results of \tool on 50 Bugs}\label{tab:rq-bug-revealing}
	\centering
    \vspace{-8pt}
	\setlength{\tabcolsep}{2pt}
\begin{tabular}{l|rr|rr|rr|rr|r}
	\toprule
	\multirow{2}{*}{Metric} & \multicolumn{2}{c|}{GPT-4o-mini} & \multicolumn{2}{c|}{Qwen3-coder-Flash} & \multicolumn{2}{c|}{Deepseek-V3.1} & \multicolumn{2}{c|}{Deepseek-V3.1-Think} & \multirow{2}{*}{\textbf{Improv.}}                                                         \\
	\cmidrule{2-3}\cmidrule{4-5}\cmidrule{6-7}\cmidrule{8-9}
	                        & Baseline                         & \tool                                  & Baseline                           & \tool                                    & Baseline                          & \tool  & Baseline & \tool                             \\
	\midrule
	Num. of Generated TCs           & {1987}                           & {{2740}}                          & {1976}                             & {2086}                                   & {2024}                            & {1797} & {2007}   & {{2661}} & -                 \\
	Pct. of Bug-revealing MTCs & 3.84\%                           & 6.53\%                                 & 4.14\%                             & 7.29\%                                   & 5.21\%                            & 6.60\% & 3.92\%   & 7.77\%        & \textbf{67.65\%} \\

	Num. of Revealed Bugs           & {4}                                & 15                                     & 5                                  & 20                                       & 7                                 & 16     & 7        & 22            & \textbf{229.46\%} \\

	\bottomrule
\end{tabular}

	\vspace{-3pt}
\end{table}

\subsubsection{Experiment Setup.}
This RQ evaluates the capability of \tool in revealing real bugs, especially for those originally discovered by metamorphic test cases.
With the collected 50 MTC-bug pairs (Section~\ref{sec:dataset}), for each bug, we take the buggy method as the target method, and measure the bug-revealing capability by the following two metrics:
\begin{itemize}[leftmargin=*]
	\item \textit{Percentage of Bug-Revealing MTCs:}
    The proportion of generated MTCs that are bug-revealing, where a bug-revealing MTC is defined as an executable MTC that fails on the buggy version but passes on the fixed version of the target method.
	\item \textit{Number of Revealed Bugs:} The number of bugs for which at least one generated MTC fails on the buggy version and passes on the fixed version.
\end{itemize}

\vspace{-10pt}
\subsubsection{Experiment Result.}
As shown in Table~\ref{tab:rq-bug-revealing},
\tool (\deepSeekThink) performed the best, successfully revealing 22 real-world bugs.
Compared with the baseline, it uncovers 15 additional bugs and improves the bug-revealing rate by 98.21\%.
When combined with other models, \tool consistently outperforms baselines, revealing 9$\sim$15 additional bugs with an average 67.65\% increase in bug-revealing MTC rate.

Based on manual inspection, we attribute the improvement to two main factors:
\begin{enumerate*}[label*=(\roman*)]
    \item \textit{Coupling-aware MR construction.}
    Some bugs are exposed only by MRs over specific method pairs, and \tool can generate such MRs through its effective coupled methods identification.
    For example, a bug~\cite{jcabiGithubCommit777a078913} in the method \code{randomRepo} is revealed when coupled with the method \code{repos}, as both invoke \code{MkRepo} and access shared fields (\code{storage}, \code{self}). This allows \tool to identify the coupling and then generate an effective MR.
    \item \textit{Input augmentation.}
    Some bugs require specific input to trigger.
    By applying additional inputs (Section~\ref{sec:app-input-generation}), \tool exercises diverse behaviors and exposes corner cases, such as boundary inputs in \code{previousClearBit} (e.g., \code{i = 1<<16})~\cite{sparseBitSetIssue13}.
\end{enumerate*}
In addition, among all evaluated LLMs, \deepSeekThink{} reveals the most bugs, likely due to its reasoning ability. By analyzing the code and understanding the coupling between methods, it can identify fault-prone interactions and construct effective MRs.

{
By analyzing the overlap of bugs revealed by \tool with different LLMs, we observed that a total of 28 bugs were revealed, with 8 detected by all models and both \deepSeekThink{} and \qwenThreeCoderFlash{} uniquely revealed two additional bugs.
This suggests that \textit{combining models could further improve the bug-revealing capability}. This is an interesting strategy for future enhancement.
Notably, 19 of the 22 bugs found by \deepSeekThink{} are exposed by multiple distinct MRs. For example, a bug in \code{multiply} can be revealed by different algebraic relations~\cite{ojAlgoIssue49}. This highlights that \textit{MR diversity plays a role in enhancing the bug-revealing capability.}
}

\textit{Failure analysis.}
While \tool successfully detected 22 (44\%) real bugs originally revealed by human-written MTCs, some bugs remained unrevealed.
A major reason is that some buggy methods are highly domain-specific and implement complex business logic, where providing only the method code is insufficient for LLMs to infer the intended behavior (e.g., a bug related to ``compaction file metrics'' in Apache IoTDB~\cite{iotdbPull13691}).
Module-level or project-level knowledge is required.
Automatically extracting and providing such knowledge as the context for LLMs remains a challenging and promising direction for future research.
In other cases, constructing inputs required access to external resources, such as environment variables or specific file contents~\cite{zinggIssue60}.
When no concrete examples were retrieved in the project under test, \tool-generated MTCs failed to set up that, resulting in non-executable or non-bug-revealing tests.
Effectively retrieving and adapting such project-level context for test generation is still an open challenge.

\begin{answertorq}
\tool{} can successfully detect 22 (out of 50) real-world bugs originally discovered by human-written MTCs.
However, some unrevealed bugs are rooted in domain-specific business logic, requiring module-level or even project-level context to construct bug-revealing MTCs. Effectively leveraging such context remains an open challenge for future work.
\end{answertorq}

\subsection{\rqablation{}: Ablation Study on \tool} \label{sec:rq-ablation}

\subsubsection{Experiment Setup.}
This RQ aims to evaluate the contribution of major steps in \tool to its overall effectiveness in generating valid and bug-revealing MTCs.
We use the same tasks and metrics as in \rqvalidity{} (validity) and \rqbugrevealing{} (bug-revealing capability).
We created three ablated variants of \tool ($v_1$, $v_2$, and $v_3$) by ablating three steps to analyze their contribution. We chose \tool built with \deepSeekThink{} which achieves the best result in \rqvalidity{} and \rqbugrevealing{} (Sections~\ref{sec:rq-validity} and \ref{sec:rq-bug-revealing}).
The variants are as follows:
\begin{itemize}[leftmargin=*]
	\item \textbf{$v_1$: \tool w/o Coupling Analysis.}
	      This variant disables the \textit{Coupling Analysis} step (Section~\ref{sec:app-relevance}), meaning no functionally coupled methods and corresponding invocation examples are provided as context to LLMs during the MTC generation.

	\item \textbf{$v_2$: \tool w/o MTC Amplification.}
	      This variant disables the \textit{{Input Augmentation}} step (Section~\ref{sec:app-input-generation}), thus no additional inputs are generated to amplify MTCs.

	\item \textbf{$v_3$: \tool w/o MTC Validation.}
	      This variant disables the \textit{Validation and Filtering} step (Section~\ref{sec:app-filtering}), meaning all generated MTCs are retained without filtering.
\end{itemize}

\subsubsection{Experiment Result.}

\begin{table}[t]
	\scriptsize
	\caption{Ablation Study on \tool (\deepSeekThink)}\label{tab:rq-ablation}
	\centering
    \vspace{-8pt}

\begin{tabular}{l|r|rrr}
	\toprule
	{Metric}                & \tool    & $v_1$: w/o Coupling Analysis    & $v_2$: w/o MTC Amplification   & $v_3$: w/o MTC Validation  \\
	\midrule
	Num. of Generated TC           & 4151     & 3367                        & 4209                       & 4367                       \\
	Pct. of Executable MTC     & 88.44\%  & 63.86\% (-27.80\%)                     & 88.69\% (0.29\%)                    & 88.99\% (0.64\%)                   \\
	Pct. of Valid MTC              & 83.57\%  & \textbf{56.28\% (-32.65\%)} & 83.61\% (0.05\%)                    & 78.84\% (-5.66\%)                    \\
	Num. of Successful Task 	& 98       & \textbf{86 (-12.24\%)}      & 98 (0.00\%)                         & 98 (0.00\%)                        \\
	Pct. of False Alarm        & 5.50\%   & \textbf{11.86 (115.53\%)}   & 5.73\% (4.13\%)                     & \textbf{9.44\% (71.56\%)} \\

	\midrule
	Pct. of Bug-revealing MTC & {7.77\%} & \textbf{3.37\% (-56.62\%)}  & \textbf{3.89\% (-49.92\%)} & {14.06\% (81.04\%)}                  \\

	Num. of Revealed Bugs       & {22}     & \textbf{12 (-45.45\%)}      & \textbf{13 (-40.91\%)}     & {24 (9.09\%)}                       \\
	\bottomrule
\end{tabular}

\end{table}

As shown in Table~\ref{tab:rq-ablation}, disabling \textit{Coupling Analysis} ($v_1$) led to a 32.65\% decrease in valid MTC rate and a 56.62\% decrease in bug-revealing rate.
This suggests that leveraging functional coupling as an explicit hint is crucial for generating valid MTCs and reducing hallucinations. This aligns with the findings in RQs of validation and bug-revealing capability.

When disabling \textit{{Input Augmentation}} ($v_2$), the number of revealed bugs significantly decreased by 40.91\% (from 22 to 13).
This highlights that applying generated MRs to additional inputs strengthens generated MTCs by exercising a wider range of program behaviors.
Nevertheless, even without input augmentation, \tool revealed more bugs compared with the baseline, indicating that MRs over multiple methods already contribute to bug revealing, and MTC amplification further boosts the bug-revealing rate by 89.94\% (from 3.89\% to 7.77\%).

{
It is observed that when disabling \textit{Input Augmentation}, \tool still successfully finished 98 tasks (i.e., generating valid MTCs for 98 target methods).
This is because \textit{Input Augmentation} is designed to apply generated MRs to more test inputs, rather than assist in generating valid MRs.
These 98 tasks were already generated with valid MTCs prior to \textit{Input Augmentation}, which then augments them with more diverse inputs.
However, without \textit{Input Augmentation}, validation based on mutation analysis becomes less effective due to reduced input diversity. As a result, more MTCs are retained (from 4,151 to 4,209), and the false-alarm rate slightly increases (from 5.50\% to 5.73\%).
}

Disabling \textit{Validation and Filtering} ($v_3$) increased the false-alarm percentage by 71.56\% (from 5.5\% to 9.44\%), indicating the effectiveness of the validation step in mitigating the false alarm issue.
The slight increase in bug-revealing rate is because some bug-revealing MTCs are filtered out together with invalid ones.
In some cases, both invalid and bug-revealing MTCs fail on both the original and mutated versions (0\% pass rate), or valid MTCs fail to kill any mutants (100\% pass rate on both), making mutation analysis based validation unable to distinguish them.
A possible mitigation is to generate more diverse inputs to improve the mutant-killing capability.

\begin{answertorq}
    Each of the three steps uniquely enhances \tool's effectiveness.
    Functional coupling helps LLMs to generate more valid MTCs, MTC amplification augments the input to reveal more bugs, and mutation analysis based validation filters nearly half of the false alarms (reducing the rate from 9.44\% to 5.5\%).
\end{answertorq}

\subsection{\rqsimilarity{}: Similarity to human-written MTCs} \label{sec:rq-similarity}

\subsubsection{Experiment Setup.}
Human-written MTCs represent well-established practices for constructing MRs, including the selection of method pairs as well as the input and output relation construction.
This RQ evaluates whether the MTCs generated by \tool{} can mimic these practices by checking if they encode the same \emph{MR-skeletons} as human-written ones.
This can demonstrate the potential of \tool to assist developers in MTCs construction, facilitating developers in integrating the generated tests into their codebase and easing subsequent maintenance.
We use the same 100 evaluation tasks as in \rqvalidity{}.

According to the definition of MTC in Section~\ref{sec:preliminaries}, an MR-skeleton consists of three core components:
\begin{enumerate*}[label=(\roman*)]
\item \textit{Input Relation:} the input transformation (e.g., API calls) applied to generate follow-up inputs, if applicable;
\item \textit{Execution:} the MR involved method pair (e.g., <\code{encryptText}, \code{decryptText}> in Listing~\ref{lst:MTC-encryptDecrypt}); and
\item \textit{Output Relation:} the assertion type (e.g., \code{assertEquals}) and the involved elements (e.g., source input, source output, follow-up output) to verify the output relation.
\end{enumerate*}
Considering that the same output relation can be implemented in multiple ways, we normalize assertions for ease of comparison.
Specifically, we normalize all assertions into comparable assertions~\cite{mrscout}. For example, boolean-style \code{assertTrue(x.equals(y))} and \code{assertFalse(x.equals(y))} are normalized to \code{assertEquals(x, y)} and \code{assertNotEquals(x, y)}, respectively. More details can be found in \tool's artifact~\cite{tool}.
Based on the definition of MR-skeleton, we take the human-written MTCs as the ground truth, and measure the similarity at two levels:
\begin{itemize}[leftmargin=*]
	\item {L1: Method-Pair consistency:} the proportion of target methods where at least one generated MTC couples the \emph{same method pair} as the human-written MTC.
	\item {L2: MR-Skeleton consistency:} the proportion of target methods where at least one generated MTC encodes the same MR-skeleton as the human-written MTC, i.e., matching the input transformation, method pair, and output relation assertion type and elements. {The MTCs satisfying L2 must satisfy L1 as well.}
\end{itemize}

\subsubsection{Experiment Result.}

\begin{table}[t]
	\scriptsize
	\caption{Similarity of \tool-Generated MTCs to Human-Written MTCs}\label{tab:rq-similar}
	\centering
	\vspace{-10pt}
	
\setlength{\tabcolsep}{1.6pt}
\begin{tabular}{l|cc|cc|cc|cc|c}
	\toprule
	\multirow{2}{*}{Metric}     & \multicolumn{2}{c|}{GPT-4o-mini} & \multicolumn{2}{c|}{Qwen3-coder-Flash} & \multicolumn{2}{c|}{Deepseek-V3.1} & \multicolumn{2}{c|}{Deepseek-V3.1-Think} & \multirow{2}{*}{\textbf{Improv.}}                                                                       \\
	\cmidrule{2-3}\cmidrule{4-5}\cmidrule{6-7}\cmidrule{8-9}
	                            & Baseline                         & \tool                                  & Baseline                           & \tool                                    & Baseline                          & \tool             & Baseline & \tool                                \\
	\midrule
	L1: Method-Pair Consistency & {61\%}                           & {89\% (+45.90\%)}                      & {61\%}                             & {86\% (+40.98\%)}                        & {74\%}                            & {87\% (+17.56\%)} & {81\%}   & {92\% (+13.58\%)} & \textbf{29.51\%} \\
	L2: MR-Skeleton Consistency & {46\%}                           & {85\% (+84.78\%)}                      & {46\%}                             & {84\% (+82.61\%)}                        & {55\%}                            & {84\% (+52.73\%)} & {65\%}   & {90\% (+38.46\%)} & \textbf{64.65\%} \\

	\bottomrule
\end{tabular}

\end{table}

Table~\ref{tab:rq-similar} shows that \tool-generated MTCs can match the human-coupled method pairs for 86$\sim$92 target methods and encode the same MR-skeletons for 84$\sim$90.
Compared to the baseline, \tool improves method-pair consistency by 29.51\% and full MR-skeleton consistency by 64.65\%.
These results highlight the effectiveness of \tool's coupling analysis.
\tool identified most functionally coupled methods used in human-written MTCs.
The high MR-skeleton consistency further demonstrates \tool's potential to assist developers in MTCs construction, integration, and maintenance.

\textit{Failure analysis.}
Despite the overall high consistency, \tool (\deepSeekThink) missed eight tasks in identifying the same method pairs and failed to encode the same MR-skeleton in ten tasks.
Our inspection revealed three main causes:
(i) some developer-selected method pairs exhibit implicit relevance in the code, such as \code{a2q} paired with \code{readAndWrite} for JSON serialization~\cite{JUnit5TestBaseGitHub};
(ii) \tool-generated MTCs sometimes construct correct but different MRs,
{e.g., for the \code{cosineSimilarity} method, the developer-constructed MR captures self-similarity dominance ($\cos(x, x) \ge \cos(x, y)$ for any vector $y \neq x$), while \tool{} derives an equivalence MR ($\cos(x, y) = \cos(x, y)$)}~\cite{SimplerPlannerTestGitHub}; and
(iii) some inconsistencies arise from equivalent but differently expressed assertions, e.g.,  \code{assertEquals(x,y)} and \code{assertTrue(x.customizedEquals(y))}~\cite{NTV2TestGitHub}.
{The concrete MTCs can be found at \tool's site~\cite{tool}.}

\begin{answertorq}
    {
	\tool can identify most of the coupled method pairs used in human-written MTCs, achieving over 90\% MR-skeleton consistency with human-written MTCs.
    This demonstrates its potential to assist developers in MTC construction.
    }
\end{answertorq}

\subsection{\rqcost{}: Cost Analysis} \label{sec:rq-cost}
To better understand the practical scalability of \tool{}, we analyze the cost of using LLMs in terms of token consumption and monetary expenditure.
We report the average cost of generating MTCs for a single target method, aggregated over all LLM invocations involved in \tool{}, including five repeated runs of MTC generation (Section~\ref{sec:evl-setup}), as well as MTC Amplification and Validation steps in \tool{}.
On average, for one target method, \tool{} costs \$0.09 (527.96k) using GPT-4o-mini, \$0.23 (501.72k) using Qwen3-coder-Flash, \$0.10 (533.02k) using Deepseek-V3.1, and \$0.40 (736.35k) using Deepseek-V3.1-Think.
Deepseek-V3.1-Think consumes 38.24\% more tokens than Deepseek-V3.1, which we attribute to additional reasoning tokens during inference. Moreover, Deepseek-V3.1-Think has a higher token price than Deepseek-V3.1, resulting in the highest cost (\$0.40 per target method). Note that our experiments were conducted in August~2025; token pricing and discount policies may change over time.

\begin{answertorq}
    \tool{} consumes around 501k\textasciitilde736k tokens (\$0.09\textasciitilde\$0.40) per target method averagely. The cost is comparable to recent LLM-based test generation approaches~\cite{CoverUp_FSE25,chen2024chatunitest}.
\end{answertorq}

\subsection{\rqmrscout{}: \tool{} V.S. MR-Scout} \label{sec:rq-vsmrscout}
To better clarify the value of \tool, we conducted a direct comparison between \tool{} and MR-Scout~\cite{mrscout} (a recent technique that discovers and synthesizes MRs from existing developer-written MTCs) from three perspectives: \textit{applicability}, \textit{correctness}, and \textit{scalability}.

\subsubsection{Applicability.}
MR-Scout suffers from two fundamental limitations in applicability.
First, it synthesized MRs from existing MTCs, and thus only works for programs (i.e., methods under test) that already have developer-written MTCs. However, MTCs are rare (1\%) in real-world projects~\cite{mrscout}, which limits its applicability.
In contrast, \tool{} infers MRs directly from the target program. It complements MR-Scout by steering away from the reliance on existing MTCs.

Second, MR-Scout requires explicit input transformations in the discovered MTCs. If such transformations are missing, MR-Scout cannot synthesize codified MRs from the discovered MTCs, and therefore, cannot generalize them to new inputs as new MTCs.
    For the 100 target methods with existing MTCs in RQ1, MR-Scout is only applicable to 21 of them, due to the absence of such transformations in the others.
In comparison, \tool{} successfully generates valid MTCs amplified with diverse inputs for 98 of the 100 target methods (Section~\ref{sec:rq-validity}).

\subsubsection{{Correctness.}}
{
For the 100 target methods in RQ1, our manual validation of the 100 sampled MTCs shows that \tool{} achieves a correctness rate of 0.89 (Section~\ref{sec:evl-correctness}).
For MR-Scout synthesized MRs, we manually validated them using the same protocol in Section~\ref{sec:evl-correctness}, and the result shows a correctness rate of 0.90.
Overall, these results suggest that \tool{} achieves comparable correctness to MR-Scout.
}

{
\textit{Differences and commonalities.}
We further analyze the overlap between MTCs constructed by \tool{} and MR-Scout.
For only {14} out of the 100 target methods, \tool{} and MR-Scout generated MTCs encoding the same MR, indicating that the overlap is minor and thus the two approaches derive complementary MRs in practice.
}

\subsubsection{Scalability.}
We compare scalability in terms of LLM cost and end-to-end runtime.
MR-Scout does not rely on LLMs and thus incurs no token or monetary cost.
\tool{} introduces additional LLM overhead, consuming 501k\textasciitilde 736k tokens (\$0.09\textasciitilde  \$0.40) per target method.
\tool{} takes 3.34 minutes per target method on average (using Deepseek-V3.1-Think). MR-Scout's runtime depends on the EvoSuite time budget (20 minutes by default) for input generation during MR validation, and its MR discovery and synthesis phases take another 20 seconds or so.

\begin{answertorq}
	{
	\tool{} and MR-Scout are complementary, as they can derive MRs under different scenarios and can be used in conjunction. \tool{} largely complements MR-Scout for the scenario that no existing MTCs are available.
	While \tool{} incurs additional LLM usage costs, it targets scenarios beyond MR-Scout's applicability and achieves comparable correctness. \tool{} enables wider adoption of metamorphic testing.
	}
\end{answertorq}

\subsection{Threats to Validity}

\textit{Representativeness of LLMs.}
Since \tool relies on LLMs for MTC and input generation, one potential threat is whether our findings based on the selected LLMs are representative.
To mitigate this threat, according to the {EvalPlus leaderboard}~\cite{urlEvalLeaderboard}, we include representative LLMs from three well-known LLM families, i.e., {\gptFourMini} from OpenAI~\cite{urlGPTFourMini},{\qwenThreeCoderFlash} from Alibaba~\cite{urlCodeQwen}, {\deepSeek} and {\deepSeekThink} from DeepSeek~\cite{urlDeepseek}.

\begin{table}[t]
	\scriptsize
	\caption{Performance of \tool (GPT-4o-mini) on 100 Target Methods Before or After the Cut-Off Date}\label{tab:rq-data-leakage-issue}
	\centering
    \vspace{-8pt}
	\setlength{\tabcolsep}{1.5pt}
\begin{tabular}{l|rrr|rr}
	\toprule
	\multirow{2}{*}{Target methods} & \multicolumn{3}{c|}{Validity} & \multicolumn{2}{c}{Similarity}                                                                              \\
	\cmidrule(l{1pt}r{1pt}){2-4}
	\cmidrule(l{1pt}r{1pt}){5-6}
	                          & {\tiny Num. of} Successful Tasks        & {\tiny Pct. of} Valid MTCs                & {\tiny Pct. of} False alarm & L1: Method-Pair consistency & L2: MR-Skeleton consistency \\
	\midrule
	Before Cut-off            & {92}                          & {71.84\%}                      & {14.16\%}        & {89}                      & {85}                        \\
	After Cut-off             & {94}                          & {73.17\%}                      & {9.52\%}         & {85}                      & {79}                        \\

	\bottomrule
\end{tabular}

\end{table}

\textit{Data Contamination.}
A potential threat is data contamination, where target programs or MTCs may have appeared in the LLMs’ pretraining data, potentially biasing results.
To mitigate this, we constructed an \emph{after cut-off} dataset following the same procedure in Section~\ref{sec:dataset}, using entries created after the GPT-4o-mini training cut-off (October 2023~\cite{gpt4omini-cufoff}).
As shown in Table~\ref{tab:rq-data-leakage-issue}, \tool achieved slightly lower similarity to human-written MTCs but a higher percentage of valid MTC compared to the \emph{before cut-off} dataset.
This indicates that \tool's effectiveness still holds for subjects after the cut-off date, and not simply an artifact of training-data memorization.

{
Another potential data leakage arises from providing underlying MTCs as examples during the MTC generation (Section~\ref{sec:app-mtc-generation}).
In the evaluation, we excluded all detected MTCs from the examples and manually inspected results for 50 buggy subjects (Section~\ref{sec:dataset-Bug-revealing MTCs}), finding no such leakage.
}

\textit{Representativeness of Experimental Subjects.}
{
A potential threat concerns the generalizability of our findings across projects and programming languages.
To mitigate the first one, we selected representative Java projects (Section~\ref{sec:dataset}), following the strategy of prior studies~\cite{mradopt,mrscout,DBLP:conf/icsm/Wang0HSX0WL20}.
Due to the substantial effort required to collect reproducible MT-revealed bugs, this study focuses on a single widely used language (Java) and evaluates the bug-revealing capability of \tool{} (RQ2) on 50 Java bugs.
This constitutes a limitation of our study --- the generalizability of our findings to other programming languages is not evaluated.
Constructing a benchmark of multiple programming languages is a meaningful direction for future work.
}

\textit{Quality of Ground Truths.}
We use human-written MTCs as ground truth and treat the fixed or latest program versions as bug-free when answering the RQs on validity, bug revealing, and similarity.
A potential threat concerns the quality of these ground truths.
To mitigate this, we applied three filtering criteria to select high-quality MTCs and corresponding target methods (Section~\ref{sec:dataset}).
{
We further manually validated the developer-written MTCs for the 100 sampled entries in RQ1, following the criteria of MR-Scout~\cite{mrscout}.
We found that 99 out of 100 MTCs are valid, with only one too complex to understand and validate.
This aligns with the 0.97 precision reported by MR-Scout~\cite{mrscout}, indicating the high quality of these ground truths.
}

\section{Discussion}
\paragraph{Type of bugs that \tool{} can detect.}
In the evaluation, the experimental bugs are known to be detectable by MTCs (Section~\ref{sec:rq-bug-revealing}).
However, \tool is not limited to bugs that are necessarily found with MTCs. This is because MTCs generated by \tool{} do not target specific types of bugs.
For example, in our experiments, Bug~\#49~\cite{examplebug49} arises from integer overflow in \code{multiply}, causing an incorrect sign flip when the numerator or denominator is large.
Given a large numerator or denominator, this bug can be revealed by a non-MR assertion (e.g., \code{assertTrue(a.multiply(b)>0)}), or an \tool{}-generated MTC over the coupled methods \code{multiply} and \code{divide} --- \code{assertEquals(a.multiply(b)/b, a.divide(b)*b)}.

\textit{Capability boundary of \tool{}.}
\tool is designed to leverage intra-class coupling for MR generation.
Specifically, it identifies functionally coupled method pairs within the same class as ingredients for MR construction.
MRs that cannot be captured by these coupling features are therefore not generated, and \tool{} currently does not support inter-class coupling.
Generating MRs over inter-class coupling presents unique challenges, such as a substantially larger search space of candidate methods and classes, and is beyond the scope of this paper. We consider it an interesting direction for future work.
\section{Related Work}
\label{sec:relatedwork}

\textit{LLM-Based Automated Test Generation.}
Recent work has demonstrated the broad utility of LLMs in software testing~\cite{terragni2025future}, including unit test generation~\cite{schafer2023empirical}, fuzzing~\cite{jiang2024fuzzing}, and test-oracle construction~\cite{bodicoat2025understanding}.
\citet{ICSE23CodaMosa,arxiv23_FDUGPT4TestGen,TSE24_CompareGPTxSBST} studied LLM-based unit test generation and reported substantial potential as well as challenges in reliability and coverage.
\citet{ICSE24_Fuzz4All} further showed that LLM prompting can drive effective, language-agnostic fuzzing.
\citet{DBLP:conf/sigsoft/HossainFDEV23} investigated neural test-oracle generation and highlighted the need for robust validation.
These advances motivate LLM-based MT, where oracle construction remains a central bottleneck.

\textit{LLM-Based MR Generation.}
Several recent studies have explored using LLMs to generate MRs and other MT artifacts.
\citet{mradopt} proposed an LLM-based technique to automatically deduce input transformations from pairs of source and follow-up inputs.
\citet{DBLP:conf/quatic/ShinPBB24} derived MRs from requirement specifications and translated them into SMRL, a domain-specific MR language, via a two-phase LLM pipeline.
\citet{DBLP:journals/infsof/ZhangCPTYZ25} proposed a human-AI hybrid MT framework that uses predefined MR patterns to generate MRs for autonomous driving systems.
\citet{DBLP:conf/iccS/TsigkanosRMK23} leveraged LLMs to discover input and output variables from scientific software documentation for constructing MRs.
\citet{hazott2025llmassisted} designed a multi-step prompt engineering pipeline to generate MRs for embedded graphics libraries, nearly doubling structural coverage and uncovering 14 new bugs compared to manually crafted MRs.
These approaches show promise but either target specific domains, depend on auxiliary artifacts (e.g., requirement documents or API manuals), or require substantial human intervention.

Some studies evaluated LLM capabilities for MR generation.
\citet{DBLP:conf/compsac/ZhangTP23} conducted a pilot study using ChatGPT~(3.5) for MR generation in autonomous driving and found it effective at proposing useful MR candidates.
\citet{luu2023chatgpt} examined ChatGPT~(GPT-4) across nine systems of varying complexity and reported that, while correct MRs can be generated, the majority of candidates were either vaguely defined or incorrect, especially for complex or previously untested systems.
\citet{DBLP:conf/saner/ZhangSLD25} evaluated MR discovery across 37 subjects and found that {4.6\textasciitilde38.6\%} of existing MRs were rediscovered but only {29.9\textasciitilde43.8\%} of generated MRs were valid.
\citet{zhang2025integrating} further showed that stronger models (e.g., GPT-4) improve MR quality over weaker ones, though human expertise remains essential for adjudication and refinement.
In contrast, \tool{} leverages LLMs' reasoning capability to derive MRs from functionally coupled methods and generate concrete executable MTCs, providing an end-to-end automated and self-validating pipeline.

\textit{Traditional MR Generation Approaches.}
Prior to LLMs, MR generation techniques relied on search-based, pattern-based, genetic-programming-based, or heuristic approaches.
\citet{genmorph,ayerdi2021generating} and \citet{gassert} generated MRs via genetic programming but assumed a regression testing scenario.
\citet{zhang2019automatic} and \citet{zhang2014search} proposed search-based approaches to inferring MRs for numeric programs.
\citet{zhou2018metamorphic} and \citet{DBLP:journals/tse/SeguraPTC18} identified MRs based on predefined patterns, limiting generalizability across domains.
\citet{nolasco2024} proposed \textsc{MemoRIA} to identify equivalence MRs from documentation.
Recently, \citet{mrscout} derived MRs from existing test cases.
However, these approaches rely on scarce resources (i.e., MR-embedded documents or tests).
{In contrast, \tool{} directly generates MTCs from the target program without relying on such resources, and is not restricted to the regression testing scenario.}
\section{Conclusion}\label{sec:conclusion}
This paper presents \tool, a fully automated approach to generate MTCs directly from the target program via functional coupling analysis.
\tool first identifies functionally coupled method pairs based on signature commonality, function call, and state interaction.
It then employs LLMs to generate concrete MTCs and refines them based on execution feedback.
Finally, \tool amplifies and then validates the MTCs based on mutation analysis.

Our evaluation shows that \tool effectively generates valid MTCs for 98\% of tasks and successfully reveals 22 confirmed bugs, improving valid MTC generation by 64.90\%, and reducing false alarms by 36.56\% compared to baselines.
Moreover, \tool-generated MTCs achieve high consistency with human-written MR-skeletons, demonstrating \tool's potential to assist or even partially replace developers in constructing effective MTCs across diverse domains.

\section{Data Availability}
\tool{} and the experimental data are available at the website~\cite{tool} and on Zenodo~\cite{toolZenodo}.



\bibliographystyle{ACM-Reference-Format}
\bibliography{src/reference}


\begin{thebibliography}{79}


\ifx \showCODEN    \undefined \def \showCODEN     #1{\unskip}     \fi
\ifx \showDOI      \undefined \def \showDOI       #1{#1}\fi
\ifx \showISBNx    \undefined \def \showISBNx     #1{\unskip}     \fi
\ifx \showISBNxiii \undefined \def \showISBNxiii  #1{\unskip}     \fi
\ifx \showISSN     \undefined \def \showISSN      #1{\unskip}     \fi
\ifx \showLCCN     \undefined \def \showLCCN      #1{\unskip}     \fi
\ifx \shownote     \undefined \def \shownote      #1{#1}          \fi
\ifx \showarticletitle \undefined \def \showarticletitle #1{#1}   \fi
\ifx \showURL      \undefined \def \showURL       {\relax}        \fi
\providecommand\bibfield[2]{#2}
\providecommand\bibinfo[2]{#2}
\providecommand\natexlab[1]{#1}
\providecommand\showeprint[2][]{arXiv:#2}

\bibitem[Alibaba(2025)]%
        {urlCodeQwen}
\bibfield{author}{\bibinfo{person}{Alibaba}.} \bibinfo{year}{2025}\natexlab{}.
\newblock \bibinfo{booktitle}{\emph{Qwen3-coder}}.
\newblock
\urldef\tempurl%
\url{https://qwenlm.github.io/blog/qwen3-coder/}
\showURL{%
Retrieved September 1, 2025 from \tempurl}


\bibitem[Allier et~al\mbox{.}(2010)]%
        {DBLP:conf/scam/AllierVDS10}
\bibfield{author}{\bibinfo{person}{Simon Allier}, \bibinfo{person}{St{\'{e}}phane Vaucher}, \bibinfo{person}{Bruno Dufour}, {and} \bibinfo{person}{Houari~A. Sahraoui}.} \bibinfo{year}{2010}\natexlab{}.
\newblock \showarticletitle{Deriving Coupling Metrics from Call Graphs}. In \bibinfo{booktitle}{\emph{Tenth {IEEE} International Working Conference on Source Code Analysis and Manipulation, {SCAM} 2010, Timisoara, Romania, 12-13 September 2010}}. \bibinfo{publisher}{{IEEE} Computer Society}, \bibinfo{pages}{43--52}.
\newblock
\urldef\tempurl%
\url{https://doi.org/10.1109/SCAM.2010.25}
\showDOI{\tempurl}


\bibitem[Altmayer~Pizzorno and Berger(2025)]%
        {CoverUp_FSE25}
\bibfield{author}{\bibinfo{person}{Juan Altmayer~Pizzorno} {and} \bibinfo{person}{Emery~D. Berger}.} \bibinfo{year}{2025}\natexlab{}.
\newblock \showarticletitle{CoverUp: Effective High Coverage Test Generation for Python}.
\newblock \bibinfo{journal}{\emph{Proc. ACM Softw. Eng.}} \bibinfo{volume}{2}, \bibinfo{number}{FSE}, Article \bibinfo{articleno}{FSE128} (\bibinfo{date}{June} \bibinfo{year}{2025}), \bibinfo{numpages}{23}~pages.
\newblock
\urldef\tempurl%
\url{https://doi.org/10.1145/3729398}
\showDOI{\tempurl}


\bibitem[Ayerdi et~al\mbox{.}(2021)]%
        {ayerdi2021generating}
\bibfield{author}{\bibinfo{person}{Jon Ayerdi}, \bibinfo{person}{Valerio Terragni}, \bibinfo{person}{Aitor Arrieta}, \bibinfo{person}{Paolo Tonella}, \bibinfo{person}{Goiuria Sagardui}, {and} \bibinfo{person}{Maite Arratibel}.} \bibinfo{year}{2021}\natexlab{}.
\newblock \showarticletitle{Generating metamorphic relations for cyber-physical systems with genetic programming: an industrial case study}. In \bibinfo{booktitle}{\emph{Joint European Software Engineering Conference and Symposium on the Foundations of Software Engineering}}. \bibinfo{publisher}{{ACM}}, \bibinfo{pages}{1264--1274}.
\newblock


\bibitem[Ayerdi et~al\mbox{.}(2024)]%
        {genmorph}
\bibfield{author}{\bibinfo{person}{Jon Ayerdi}, \bibinfo{person}{Valerio Terragni}, \bibinfo{person}{Gunel Jahangirova}, \bibinfo{person}{Aitor Arrieta}, {and} \bibinfo{person}{Paolo Tonella}.} \bibinfo{year}{2024}\natexlab{}.
\newblock \showarticletitle{GenMorph: Automatically Generating Metamorphic Relations via Genetic Programming}.
\newblock \bibinfo{journal}{\emph{IEEE Transactions on Software Engineering}} (\bibinfo{year}{2024}), \bibinfo{pages}{1--12}.
\newblock


\bibitem[Blasi et~al\mbox{.}(2021)]%
        {DBLP:journals/jss/BlasiGEPC21}
\bibfield{author}{\bibinfo{person}{Arianna Blasi}, \bibinfo{person}{Alessandra Gorla}, \bibinfo{person}{Michael~D. Ernst}, \bibinfo{person}{Mauro Pezz{\`{e}}}, {and} \bibinfo{person}{Antonio Carzaniga}.} \bibinfo{year}{2021}\natexlab{}.
\newblock \showarticletitle{MeMo: Automatically identifying metamorphic relations in Javadoc comments for test automation}.
\newblock \bibinfo{journal}{\emph{J. Syst. Softw.}}  \bibinfo{volume}{181} (\bibinfo{year}{2021}), \bibinfo{pages}{111041}.
\newblock
\urldef\tempurl%
\url{https://doi.org/10.1016/J.JSS.2021.111041}
\showDOI{\tempurl}


\bibitem[Bodicoat et~al\mbox{.}(2025)]%
        {bodicoat2025understanding}
\bibfield{author}{\bibinfo{person}{Adam Bodicoat}, \bibinfo{person}{Gunel Jahangirova}, {and} \bibinfo{person}{Valerio Terragni}.} \bibinfo{year}{2025}\natexlab{}.
\newblock \showarticletitle{Understanding LLM-Driven Test Oracle Generation}. In \bibinfo{booktitle}{\emph{2025 2nd IEEE/ACM International Conference on AI-powered Software (AIware)}}. IEEE, \bibinfo{pages}{29--39}.
\newblock


\bibitem[Cao et~al\mbox{.}(2022)]%
        {jialunMTtranslator}
\bibfield{author}{\bibinfo{person}{Jialun Cao}, \bibinfo{person}{Meiziniu Li}, \bibinfo{person}{Yeting Li}, \bibinfo{person}{Ming Wen}, \bibinfo{person}{Shing{-}Chi Cheung}, {and} \bibinfo{person}{Haiming Chen}.} \bibinfo{year}{2022}\natexlab{}.
\newblock \showarticletitle{SemMT: {A} Semantic-Based Testing Approach for Machine Translation Systems}.
\newblock \bibinfo{journal}{\emph{ACM Transactions on Software Engineering and Methodology}} \bibinfo{volume}{31}, \bibinfo{number}{2} (\bibinfo{year}{2022}), \bibinfo{pages}{34e:1--34e:36}.
\newblock


\bibitem[Cao et~al\mbox{.}(2024)]%
        {DBLP:journals/corr/abs-2403-16898}
\bibfield{author}{\bibinfo{person}{Jialun Cao}, \bibinfo{person}{Wuqi Zhang}, {and} \bibinfo{person}{Shing{-}Chi Cheung}.} \bibinfo{year}{2024}\natexlab{}.
\newblock \showarticletitle{Concerned with Data Contamination? Assessing Countermeasures in Code Language Model}.
\newblock \bibinfo{journal}{\emph{CoRR}}  \bibinfo{volume}{abs/2403.16898} (\bibinfo{year}{2024}).
\newblock
\showeprint[arXiv]{2403.16898}


\bibitem[Chen et~al\mbox{.}(2021)]%
        {DBLP:conf/kbse/ChenJX21}
\bibfield{author}{\bibinfo{person}{Songqiang Chen}, \bibinfo{person}{Shuo Jin}, {and} \bibinfo{person}{Xiaoyuan Xie}.} \bibinfo{year}{2021}\natexlab{}.
\newblock \showarticletitle{Testing Your Question Answering Software via Asking Recursively}. In \bibinfo{booktitle}{\emph{International Conference on Automated Software Engineering}}. \bibinfo{publisher}{{IEEE}}, \bibinfo{pages}{104--116}.
\newblock


\bibitem[Chen et~al\mbox{.}(2018)]%
        {chen2018survey}
\bibfield{author}{\bibinfo{person}{Tsong~Yueh Chen}, \bibinfo{person}{Fei{-}Ching Kuo}, \bibinfo{person}{Huai Liu}, \bibinfo{person}{Pak{-}Lok Poon}, \bibinfo{person}{Dave Towey}, \bibinfo{person}{T.~H. Tse}, {and} \bibinfo{person}{Zhi~Quan Zhou}.} \bibinfo{year}{2018}\natexlab{}.
\newblock \showarticletitle{Metamorphic Testing: {A} Review of Challenges and Opportunities}.
\newblock \bibinfo{journal}{\emph{{ACM} Comput. Surv.}} \bibinfo{volume}{51}, \bibinfo{number}{1} (\bibinfo{year}{2018}), \bibinfo{pages}{4:1--4:27}.
\newblock
\urldef\tempurl%
\url{https://doi.org/10.1145/3143561}
\showDOI{\tempurl}


\bibitem[Chen et~al\mbox{.}(2016)]%
        {chen2016metric}
\bibfield{author}{\bibinfo{person}{Tsong~Yueh Chen}, \bibinfo{person}{Pak{-}Lok Poon}, {and} \bibinfo{person}{Xiaoyuan Xie}.} \bibinfo{year}{2016}\natexlab{}.
\newblock \showarticletitle{{METRIC:} METamorphic Relation Identification based on the Category-choice framework}.
\newblock \bibinfo{journal}{\emph{J. Syst. Softw.}}  \bibinfo{volume}{116} (\bibinfo{year}{2016}), \bibinfo{pages}{177--190}.
\newblock
\urldef\tempurl%
\url{https://doi.org/10.1016/j.jss.2015.07.037}
\showDOI{\tempurl}


\bibitem[Chen et~al\mbox{.}(2024)]%
        {chen2024chatunitest}
\bibfield{author}{\bibinfo{person}{Yinghao Chen}, \bibinfo{person}{Zehao Hu}, \bibinfo{person}{Chen Zhi}, \bibinfo{person}{Junxiao Han}, \bibinfo{person}{Shuiguang Deng}, {and} \bibinfo{person}{Jianwei Yin}.} \bibinfo{year}{2024}\natexlab{}.
\newblock \showarticletitle{ChatUniTest: A Framework for LLM-Based Test Generation}. In \bibinfo{booktitle}{\emph{Companion Proceedings of the 32nd ACM International Conference on the Foundations of Software Engineering}}. \bibinfo{pages}{572--576}.
\newblock


\bibitem[Cho et~al\mbox{.}(2025a)]%
        {cho2025llmorph}
\bibfield{author}{\bibinfo{person}{Steven Cho}, \bibinfo{person}{Stefano Ruberto}, {and} \bibinfo{person}{Valerio Terragni}.} \bibinfo{year}{2025}\natexlab{a}.
\newblock \showarticletitle{{LLMORPH}: Automated Metamorphic Testing of Large Language Models}. In \bibinfo{booktitle}{\emph{Proceedings of the 40th IEEE/ACM International Conference on Automated Software Engineering}}. \bibinfo{pages}{4102–4105}.
\newblock
\urldef\tempurl%
\url{https://doi.org/10.1109/ASE63991.2025.00385}
\showDOI{\tempurl}


\bibitem[Cho et~al\mbox{.}(2025b)]%
        {cho2025metamorphic}
\bibfield{author}{\bibinfo{person}{Steven Cho}, \bibinfo{person}{Stefano Ruberto}, {and} \bibinfo{person}{Valerio Terragni}.} \bibinfo{year}{2025}\natexlab{b}.
\newblock \showarticletitle{Metamorphic Testing of Large Language Models for Natural Language Processing}. In \bibinfo{booktitle}{\emph{Proceedings of the 41st IEEE International Conference on Software Maintenance and Evolution (ICSME)}}. \bibinfo{publisher}{IEEE}, \bibinfo{pages}{174--186}.
\newblock
\urldef\tempurl%
\url{https://doi.org/10.1109/ICSME64153.2025.00025}
\showDOI{\tempurl}


\bibitem[DeepSeek(2025)]%
        {urlDeepseek}
\bibfield{author}{\bibinfo{person}{DeepSeek}.} \bibinfo{year}{2025}\natexlab{}.
\newblock \bibinfo{booktitle}{\emph{DeepSeek-V3.1}}.
\newblock
\urldef\tempurl%
\url{https://api-docs.deepseek.com/news/news250821}
\showURL{%
Retrieved September 1, 2025 from \tempurl}


\bibitem[Diennea(2025)]%
        {SimplerPlannerTestGitHub}
\bibfield{author}{\bibinfo{person}{Diennea}.} \bibinfo{year}{2025}\natexlab{}.
\newblock \bibinfo{booktitle}{\emph{SimplerPlannerTest}}.
\newblock
\urldef\tempurl%
\url{https://github.com/diennea/herddb/blob/master/herddb-core/src/test/java/herddb/sql/SimplerPlannerTest.java}
\showURL{%
\tempurl}


\bibitem[Du et~al\mbox{.}(2024)]%
        {DBLP:conf/icse/Du0WWL0FS0L24}
\bibfield{author}{\bibinfo{person}{Xueying Du}, \bibinfo{person}{Mingwei Liu}, \bibinfo{person}{Kaixin Wang}, \bibinfo{person}{Hanlin Wang}, \bibinfo{person}{Junwei Liu}, \bibinfo{person}{Yixuan Chen}, \bibinfo{person}{Jiayi Feng}, \bibinfo{person}{Chaofeng Sha}, \bibinfo{person}{Xin Peng}, {and} \bibinfo{person}{Yiling Lou}.} \bibinfo{year}{2024}\natexlab{}.
\newblock \showarticletitle{Evaluating Large Language Models in Class-Level Code Generation}. In \bibinfo{booktitle}{\emph{nternational Conference on Software Engineering}}. \bibinfo{publisher}{{ACM}}, \bibinfo{pages}{81:1--81:13}.
\newblock


\bibitem[Eghbali and Pradel(2024)]%
        {DBLP:journals/corr/abs-2401-01701}
\bibfield{author}{\bibinfo{person}{Aryaz Eghbali} {and} \bibinfo{person}{Michael Pradel}.} \bibinfo{year}{2024}\natexlab{}.
\newblock \showarticletitle{De-Hallucinator: Iterative Grounding for LLM-Based Code Completion}.
\newblock \bibinfo{journal}{\emph{CoRR}}  \bibinfo{volume}{abs/2401.01701} (\bibinfo{year}{2024}).
\newblock
\showeprint[arXiv]{2401.01701}


\bibitem[Evalplus(2025)]%
        {urlEvalLeaderboard}
\bibfield{author}{\bibinfo{person}{Evalplus}.} \bibinfo{year}{2025}\natexlab{}.
\newblock \bibinfo{booktitle}{\emph{leaderboard}}.
\newblock
\urldef\tempurl%
\url{https://evalplus.github.io/leaderboard.html}
\showURL{%
Retrieved September 1, 2025 from \tempurl}


\bibitem[FasterXML(2025)]%
        {JUnit5TestBaseGitHub}
\bibfield{author}{\bibinfo{person}{FasterXML}.} \bibinfo{year}{2025}\natexlab{}.
\newblock \bibinfo{booktitle}{\emph{BasicParserFilteringTest}}.
\newblock
\urldef\tempurl%
\url{https://github.com/FasterXML/jackson-core/blob/3.x/src/test/java/tools/jackson/core/unittest/filter/BasicParserFilteringTest.java#L432}
\showURL{%
\tempurl}


\bibitem[Fregnan et~al\mbox{.}(2019)]%
        {DBLP:journals/infsof/FregnanBPB19}
\bibfield{author}{\bibinfo{person}{Enrico Fregnan}, \bibinfo{person}{Tobias Baum}, \bibinfo{person}{Fabio Palomba}, {and} \bibinfo{person}{Alberto Bacchelli}.} \bibinfo{year}{2019}\natexlab{}.
\newblock \showarticletitle{A survey on software coupling relations and tools}.
\newblock \bibinfo{journal}{\emph{Inf. Softw. Technol.}}  \bibinfo{volume}{107} (\bibinfo{year}{2019}), \bibinfo{pages}{159--178}.
\newblock
\urldef\tempurl%
\url{https://doi.org/10.1016/J.INFSOF.2018.11.008}
\showDOI{\tempurl}


\bibitem[Hazott and Gro{\ss}e(2025)]%
        {hazott2025llmassisted}
\bibfield{author}{\bibinfo{person}{Christoph Hazott} {and} \bibinfo{person}{Daniel Gro{\ss}e}.} \bibinfo{year}{2025}\natexlab{}.
\newblock \showarticletitle{{LLM}-assisted Metamorphic Testing of Embedded Graphics Libraries}. In \bibinfo{booktitle}{\emph{Forum on Specification and Design Languages}}.
\newblock
\urldef\tempurl%
\url{https://ics.jku.at/files/2025FDL_LLM-assisted_Metamorphic_Testing_of_Embedded_Graphics_Libraries.pdf}
\showURL{%
\tempurl}


\bibitem[Hossain et~al\mbox{.}(2023)]%
        {DBLP:conf/sigsoft/HossainFDEV23}
\bibfield{author}{\bibinfo{person}{Soneya~Binta Hossain}, \bibinfo{person}{Antonio Filieri}, \bibinfo{person}{Matthew~B. Dwyer}, \bibinfo{person}{Sebastian~G. Elbaum}, {and} \bibinfo{person}{Willem Visser}.} \bibinfo{year}{2023}\natexlab{}.
\newblock \showarticletitle{Neural-Based Test Oracle Generation: {A} Large-Scale Evaluation and Lessons Learned}. In \bibinfo{booktitle}{\emph{Proceedings of the 31st {ACM} Joint European Software Engineering Conference and Symposium on the Foundations of Software Engineering, {ESEC/FSE} 2023, San Francisco, CA, USA, December 3-9, 2023}}, \bibfield{editor}{\bibinfo{person}{Satish Chandra}, \bibinfo{person}{Kelly Blincoe}, {and} \bibinfo{person}{Paolo Tonella}} (Eds.). \bibinfo{publisher}{{ACM}}, \bibinfo{pages}{120--132}.
\newblock
\urldef\tempurl%
\url{https://doi.org/10.1145/3611643.3616265}
\showDOI{\tempurl}


\bibitem[IoTDB(2025)]%
        {iotdbPull13691}
\bibfield{author}{\bibinfo{person}{Apache IoTDB}.} \bibinfo{year}{2025}\natexlab{}.
\newblock \bibinfo{booktitle}{\emph{IoTDB Issue \#13691}}.
\newblock
\urldef\tempurl%
\url{https://github.com/apache/iotdb/pull/13691}
\showURL{%
\tempurl}


\bibitem[JavaParser(2024)]%
        {urlJavaParser}
\bibfield{author}{\bibinfo{person}{JavaParser}.} \bibinfo{year}{2024}\natexlab{}.
\newblock \bibinfo{booktitle}{\emph{JavaParser}}.
\newblock
\urldef\tempurl%
\url{https://javaparser.org/}
\showURL{%
Retrieved June 6, 2024 from \tempurl}


\bibitem[Jcabi(2025)]%
        {jcabiGithubCommit777a078913}
\bibfield{author}{\bibinfo{person}{Jcabi}.} \bibinfo{year}{2025}\natexlab{}.
\newblock \bibinfo{booktitle}{\emph{GitHub Commit 777a078913}}.
\newblock
\urldef\tempurl%
\url{https://github.com/jcabi/jcabi-github/commit/777a078913}
\showURL{%
\tempurl}


\bibitem[Jiang et~al\mbox{.}(2024)]%
        {jiang2024fuzzing}
\bibfield{author}{\bibinfo{person}{Yu Jiang}, \bibinfo{person}{Jie Liang}, \bibinfo{person}{Fuchen Ma}, \bibinfo{person}{Yuanliang Chen}, \bibinfo{person}{Chijin Zhou}, \bibinfo{person}{Yuheng Shen}, \bibinfo{person}{Zhiyong Wu}, \bibinfo{person}{Jingzhou Fu}, \bibinfo{person}{Mingzhe Wang}, \bibinfo{person}{Shanshan Li}, {et~al\mbox{.}}} \bibinfo{year}{2024}\natexlab{}.
\newblock \showarticletitle{When fuzzing meets llms: Challenges and opportunities}. In \bibinfo{booktitle}{\emph{Companion Proceedings of the 32nd ACM International Conference on the Foundations of Software Engineering}}. \bibinfo{pages}{492--496}.
\newblock


\bibitem[{Knowledge Cutoff Information of GPT-4o-mini}({[n.\,d.]})]%
        {gpt4omini-cufoff}
{Knowledge Cutoff Information of GPT-4o-mini} \bibinfo{year}{[n.\,d.]}\natexlab{}.
\newblock
\newblock
\urldef\tempurl%
\url{https://community.openai.com/t/introducing-gpt-4o-mini-in-the-api/871594}
\showURL{%
\tempurl}


\bibitem[Le et~al\mbox{.}(2014)]%
        {MT4Compiler_PLDI14SZD}
\bibfield{author}{\bibinfo{person}{Vu Le}, \bibinfo{person}{Mehrdad Afshari}, {and} \bibinfo{person}{Zhendong Su}.} \bibinfo{year}{2014}\natexlab{}.
\newblock \showarticletitle{Compiler validation via equivalence modulo inputs}. In \bibinfo{booktitle}{\emph{Conference on Programming Language Design and Implementation}}. \bibinfo{publisher}{{ACM}}, \bibinfo{pages}{216--226}.
\newblock


\bibitem[Lemieux et~al\mbox{.}(2023)]%
        {ICSE23CodaMosa}
\bibfield{author}{\bibinfo{person}{Caroline Lemieux}, \bibinfo{person}{Jeevana~Priya Inala}, \bibinfo{person}{Shuvendu~K. Lahiri}, {and} \bibinfo{person}{Siddhartha Sen}.} \bibinfo{year}{2023}\natexlab{}.
\newblock \showarticletitle{CodaMosa: Escaping Coverage Plateaus in Test Generation with Pre-trained Large Language Models}. In \bibinfo{booktitle}{\emph{International Conference on Software Engineering}}. \bibinfo{publisher}{{IEEE}}, \bibinfo{pages}{919--931}.
\newblock


\bibitem[Li et~al\mbox{.}(2023)]%
        {ChainOfCode}
\bibfield{author}{\bibinfo{person}{Chengshu Li}, \bibinfo{person}{Jacky Liang}, \bibinfo{person}{Andy Zeng}, \bibinfo{person}{Xinyun Chen}, \bibinfo{person}{Karol Hausman}, \bibinfo{person}{Dorsa Sadigh}, \bibinfo{person}{Sergey Levine}, \bibinfo{person}{Li Fei{-}Fei}, \bibinfo{person}{Fei Xia}, {and} \bibinfo{person}{Brian Ichter}.} \bibinfo{year}{2023}\natexlab{}.
\newblock \showarticletitle{Chain of Code: Reasoning with a Language Model-Augmented Code Emulator}.
\newblock \bibinfo{journal}{\emph{CoRR}}  \bibinfo{volume}{abs/2312.04474} (\bibinfo{year}{2023}).
\newblock
\showeprint[arXiv]{2312.04474}


\bibitem[Li et~al\mbox{.}(2025a)]%
        {li2025mdpmorpi}
\bibfield{author}{\bibinfo{person}{Jiapeng Li}, \bibinfo{person}{Zheng Zheng}, \bibinfo{person}{Yuning Xing}, \bibinfo{person}{Daixu Ren}, \bibinfo{person}{Steven Cho}, {and} \bibinfo{person}{Valerio Terragni}.} \bibinfo{year}{2025}\natexlab{a}.
\newblock \showarticletitle{MDPMORPH: An MDP-Based Metamorphic Testing Framework for Deep Reinforcement Learning Agents}. In \bibinfo{booktitle}{\emph{Proceedings of the 36th IEEE International Symposium on Software Reliability Engineering}}. \bibinfo{pages}{154--166}.
\newblock
\urldef\tempurl%
\url{https://doi.org/10.1109/ISSRE66568.2025.00028}
\showDOI{\tempurl}


\bibitem[Li et~al\mbox{.}(2025b)]%
        {li2025mdpmorph}
\bibfield{author}{\bibinfo{person}{Jiapeng Li}, \bibinfo{person}{Zheng Zheng}, \bibinfo{person}{Yuning Xing}, \bibinfo{person}{Daixu Ren}, \bibinfo{person}{Steven Cho}, {and} \bibinfo{person}{Valerio Terragni}.} \bibinfo{year}{2025}\natexlab{b}.
\newblock \showarticletitle{Metamorphic Testing of Deep Reinforcement Learning Agents with {MDPMORPH}}. In \bibinfo{booktitle}{\emph{Proceedings of the 40th IEEE/ACM International Conference on Automated Software Engineering}}. \bibinfo{pages}{4086--4089}.
\newblock
\urldef\tempurl%
\url{https://doi.org/10.1109/ASE63991.2025.00381}
\showDOI{\tempurl}


\bibitem[Li et~al\mbox{.}(2022)]%
        {alphacode}
\bibfield{author}{\bibinfo{person}{Yujia Li}, \bibinfo{person}{David~H. Choi}, \bibinfo{person}{Junyoung Chung}, \bibinfo{person}{Nate Kushman}, \bibinfo{person}{Julian Schrittwieser}, {and} \bibinfo{person}{et al.}} \bibinfo{year}{2022}\natexlab{}.
\newblock \showarticletitle{Competition-Level Code Generation with AlphaCode}.
\newblock \bibinfo{journal}{\emph{CoRR}}  \bibinfo{volume}{abs/2203.07814} (\bibinfo{year}{2022}).
\newblock
\showeprint[arXiv]{2203.07814}


\bibitem[Liu et~al\mbox{.}(2014)]%
        {DBLP:journals/tse/LiuKTC14}
\bibfield{author}{\bibinfo{person}{Huai Liu}, \bibinfo{person}{Fei{-}Ching Kuo}, \bibinfo{person}{Dave Towey}, {and} \bibinfo{person}{Tsong~Yueh Chen}.} \bibinfo{year}{2014}\natexlab{}.
\newblock \showarticletitle{How Effectively Does Metamorphic Testing Alleviate the Oracle Problem?}
\newblock \bibinfo{journal}{\emph{IEEE Transactions on Software Engineering}} \bibinfo{volume}{40}, \bibinfo{number}{1} (\bibinfo{year}{2014}), \bibinfo{pages}{4--22}.
\newblock


\bibitem[LocationTech(2025)]%
        {NTV2TestGitHub}
\bibfield{author}{\bibinfo{person}{LocationTech}.} \bibinfo{year}{2025}\natexlab{}.
\newblock \bibinfo{booktitle}{\emph{NTV2Test}}.
\newblock
\urldef\tempurl%
\url{https://github.com/locationtech/proj4j/blob/master/core/src/test/java/org/locationtech/proj4j/datum/NTV2Test.java}
\showURL{%
\tempurl}


\bibitem[Luu et~al\mbox{.}(2023)]%
        {luu2023chatgpt}
\bibfield{author}{\bibinfo{person}{Quang{-}Hung Luu}, \bibinfo{person}{Huai Liu}, {and} \bibinfo{person}{Tsong~Yueh Chen}.} \bibinfo{year}{2023}\natexlab{}.
\newblock \showarticletitle{Can ChatGPT Advance Software Testing Intelligence? {A}n Experience Report on Metamorphic Testing}.
\newblock \bibinfo{journal}{\emph{CoRR}}  \bibinfo{volume}{abs/2310.19204} (\bibinfo{year}{2023}).
\newblock
\showeprint[arXiv]{2310.19204}
\urldef\tempurl%
\url{https://arxiv.org/abs/2310.19204}
\showURL{%
\tempurl}


\bibitem[Ma et~al\mbox{.}(2023)]%
        {DBLP:conf/issta/MaS00C23}
\bibfield{author}{\bibinfo{person}{Haoyang Ma}, \bibinfo{person}{Qingchao Shen}, \bibinfo{person}{Yongqiang Tian}, \bibinfo{person}{Junjie Chen}, {and} \bibinfo{person}{Shing{-}Chi Cheung}.} \bibinfo{year}{2023}\natexlab{}.
\newblock \showarticletitle{Fuzzing Deep Learning Compilers with HirGen}. In \bibinfo{booktitle}{\emph{International Symposium on Software Testing and Analysis}}. \bibinfo{publisher}{{ACM}}, \bibinfo{pages}{248--260}.
\newblock


\bibitem[Major(2025)]%
        {major}
\bibfield{author}{\bibinfo{person}{Major}.} \bibinfo{year}{2025}\natexlab{}.
\newblock \bibinfo{booktitle}{\emph{Mutation Testing}}.
\newblock
\urldef\tempurl%
\url{https://mutation-testing.org/}
\showURL{%
\tempurl}


\bibitem[Nolasco et~al\mbox{.}(2024)]%
        {nolasco2024}
\bibfield{author}{\bibinfo{person}{Agust{\'{\i}}n Nolasco}, \bibinfo{person}{Facundo Molina}, \bibinfo{person}{Renzo Degiovanni}, \bibinfo{person}{Alessandra Gorla}, \bibinfo{person}{Diego Garbervetsky}, \bibinfo{person}{Mike Papadakis}, \bibinfo{person}{Sebasti{\'{a}}n Uchitel}, \bibinfo{person}{Nazareno Aguirre}, {and} \bibinfo{person}{Marcelo~F. Frias}.} \bibinfo{year}{2024}\natexlab{}.
\newblock \showarticletitle{Abstraction-Aware Inference of Metamorphic Relations}.
\newblock \bibinfo{journal}{\emph{Proceedings of the ACM on Software Engineering}} \bibinfo{volume}{1}, \bibinfo{number}{{FSE}} (\bibinfo{year}{2024}), \bibinfo{pages}{450--472}.
\newblock


\bibitem[OpenAI(2025)]%
        {urlGPTFourMini}
\bibfield{author}{\bibinfo{person}{OpenAI}.} \bibinfo{year}{2025}\natexlab{}.
\newblock \bibinfo{booktitle}{\emph{GPT-4o mini}}.
\newblock
\urldef\tempurl%
\url{https://platform.openai.com/docs/models/gpt-4o-mini}
\showURL{%
Retrieved September 1, 2025 from \tempurl}


\bibitem[Optimatika(2025a)]%
        {ojAlgoIssue49}
\bibfield{author}{\bibinfo{person}{Optimatika}.} \bibinfo{year}{2025}\natexlab{a}.
\newblock \bibinfo{booktitle}{\emph{OjAlgo Issue \#49}}.
\newblock
\urldef\tempurl%
\url{https://github.com/optimatika/ojAlgo/issues/49}
\showURL{%
\tempurl}


\bibitem[Optimatika(2025b)]%
        {examplebug49}
\bibfield{author}{\bibinfo{person}{Optimatika}.} \bibinfo{year}{2025}\natexlab{b}.
\newblock \bibinfo{booktitle}{\emph{OjAlgo Issue \#49}}.
\newblock
\urldef\tempurl%
\url{https://github.com/optimatika/ojAlgo/issues/49}
\showURL{%
Retrieved September 1, 2025 from \tempurl}


\bibitem[Poshyvanyk et~al\mbox{.}(2009)]%
        {DBLP:journals/ese/PoshyvanykMFG09}
\bibfield{author}{\bibinfo{person}{Denys Poshyvanyk}, \bibinfo{person}{Andrian Marcus}, \bibinfo{person}{Rudolf Ferenc}, {and} \bibinfo{person}{Tibor Gyim{\'{o}}thy}.} \bibinfo{year}{2009}\natexlab{}.
\newblock \showarticletitle{Using information retrieval based coupling measures for impact analysis}.
\newblock \bibinfo{journal}{\emph{Empir. Softw. Eng.}} \bibinfo{volume}{14}, \bibinfo{number}{1} (\bibinfo{year}{2009}), \bibinfo{pages}{5--32}.
\newblock
\urldef\tempurl%
\url{https://doi.org/10.1007/S10664-008-9088-2}
\showDOI{\tempurl}


\bibitem[Ravi et~al\mbox{.}(2025)]%
        {ravi2025llmloop}
\bibfield{author}{\bibinfo{person}{Ravin Ravi}, \bibinfo{person}{Dylan Bradshaw}, \bibinfo{person}{Stefano Ruberto}, \bibinfo{person}{Gunel Jahangirova}, {and} \bibinfo{person}{Valerio Terragni}.} \bibinfo{year}{2025}\natexlab{}.
\newblock \showarticletitle{LLMLOOP: Improving LLM-Generated Code and Tests through Automated Iterative Feedback Loops}. In \bibinfo{booktitle}{\emph{2025 IEEE International Conference on Software Maintenance and Evolution (ICSME)}}. IEEE, \bibinfo{pages}{930--934}.
\newblock


\bibitem[Sch{\"a}fer et~al\mbox{.}(2023)]%
        {schafer2023empirical}
\bibfield{author}{\bibinfo{person}{Max Sch{\"a}fer}, \bibinfo{person}{Sarah Nadi}, \bibinfo{person}{Aryaz Eghbali}, {and} \bibinfo{person}{Frank Tip}.} \bibinfo{year}{2023}\natexlab{}.
\newblock \showarticletitle{An empirical evaluation of using large language models for automated unit test generation}.
\newblock \bibinfo{journal}{\emph{IEEE Transactions on Software Engineering}} \bibinfo{volume}{50}, \bibinfo{number}{1} (\bibinfo{year}{2023}), \bibinfo{pages}{85--105}.
\newblock


\bibitem[Segura et~al\mbox{.}(2016)]%
        {segura2016survey}
\bibfield{author}{\bibinfo{person}{Sergio Segura}, \bibinfo{person}{Gordon Fraser}, \bibinfo{person}{Ana~Bel{\'{e}}n S{\'{a}}nchez}, {and} \bibinfo{person}{Antonio~Ruiz Cort{\'{e}}s}.} \bibinfo{year}{2016}\natexlab{}.
\newblock \showarticletitle{A Survey on Metamorphic Testing}.
\newblock \bibinfo{journal}{\emph{{IEEE} Trans. Software Eng.}} \bibinfo{volume}{42}, \bibinfo{number}{9} (\bibinfo{year}{2016}), \bibinfo{pages}{805--824}.
\newblock
\urldef\tempurl%
\url{https://doi.org/10.1109/TSE.2016.2532875}
\showDOI{\tempurl}


\bibitem[Segura et~al\mbox{.}(2018)]%
        {DBLP:journals/tse/SeguraPTC18}
\bibfield{author}{\bibinfo{person}{Sergio Segura}, \bibinfo{person}{Jos{\'{e}}~Antonio Parejo}, \bibinfo{person}{Javier Troya}, {and} \bibinfo{person}{Antonio~Ruiz Cort{\'{e}}s}.} \bibinfo{year}{2018}\natexlab{}.
\newblock \showarticletitle{Metamorphic Testing of RESTful Web APIs}.
\newblock \bibinfo{journal}{\emph{IEEE Transactions on Software Engineering}} \bibinfo{volume}{44}, \bibinfo{number}{11} (\bibinfo{year}{2018}), \bibinfo{pages}{1083--1099}.
\newblock


\bibitem[Shin et~al\mbox{.}(2024)]%
        {DBLP:conf/quatic/ShinPBB24}
\bibfield{author}{\bibinfo{person}{Seung~Yeob Shin}, \bibinfo{person}{Fabrizio Pastore}, \bibinfo{person}{Domenico Bianculli}, {and} \bibinfo{person}{Alexandra Baicoianu}.} \bibinfo{year}{2024}\natexlab{}.
\newblock \showarticletitle{Towards Generating Executable Metamorphic Relations Using Large Language Models}. In \bibinfo{booktitle}{\emph{Quality of Information and Communications Technology - 17th International Conference on the Quality of Information and Communications Technology, {QUATIC} 2024, Pisa, Italy, September 11-13, 2024, Proceedings}} \emph{(\bibinfo{series}{Communications in Computer and Information Science}, Vol.~\bibinfo{volume}{2178})}, \bibfield{editor}{\bibinfo{person}{Antonia Bertolino}, \bibinfo{person}{Jo{\~{a}}o~Pascoal Faria}, \bibinfo{person}{Patricia Lago}, {and} \bibinfo{person}{Laura Semini}} (Eds.). \bibinfo{publisher}{Springer}, \bibinfo{pages}{126--141}.
\newblock
\urldef\tempurl%
\url{https://doi.org/10.1007/978-3-031-70245-7\_9}
\showDOI{\tempurl}


\bibitem[Sun et~al\mbox{.}(2016a)]%
        {MT4Compiler_OOPSLA16SZD}
\bibfield{author}{\bibinfo{person}{Chengnian Sun}, \bibinfo{person}{Vu Le}, {and} \bibinfo{person}{Zhendong Su}.} \bibinfo{year}{2016}\natexlab{a}.
\newblock \showarticletitle{Finding compiler bugs via live code mutation}. In \bibinfo{booktitle}{\emph{International Conference on Object-Oriented Programming, Systems, Languages, and Applications,}}. \bibinfo{publisher}{{ACM}}, \bibinfo{pages}{849--863}.
\newblock


\bibitem[Sun et~al\mbox{.}(2016b)]%
        {sun2016mumt}
\bibfield{author}{\bibinfo{person}{Chang{-}Ai Sun}, \bibinfo{person}{Yiqiang Liu}, \bibinfo{person}{Zuoyi Wang}, {and} \bibinfo{person}{W.~K. Chan}.} \bibinfo{year}{2016}\natexlab{b}.
\newblock \showarticletitle{{\(\mu\)}MT: a data mutation directed metamorphic relation acquisition methodology}. In \bibinfo{booktitle}{\emph{International Workshop on Metamorphic Testing}}. \bibinfo{publisher}{{ACM}}, \bibinfo{pages}{12--18}.
\newblock


\bibitem[Tang et~al\mbox{.}(2024)]%
        {TSE24_CompareGPTxSBST}
\bibfield{author}{\bibinfo{person}{Yutian Tang}, \bibinfo{person}{Zhijie Liu}, \bibinfo{person}{Zhichao Zhou}, {and} \bibinfo{person}{Xiapu Luo}.} \bibinfo{year}{2024}\natexlab{}.
\newblock \showarticletitle{ChatGPT vs SBST: A Comparative Assessment of Unit Test Suite Generation}.
\newblock \bibinfo{journal}{\emph{IEEE Transactions on Software Engineering}} (\bibinfo{year}{2024}), \bibinfo{pages}{1--19}.
\newblock


\bibitem[Terragni et~al\mbox{.}(2020)]%
        {gassert}
\bibfield{author}{\bibinfo{person}{Valerio Terragni}, \bibinfo{person}{Gunel Jahangirova}, \bibinfo{person}{Paolo Tonella}, {and} \bibinfo{person}{Mauro Pezzè}.} \bibinfo{year}{2020}\natexlab{}.
\newblock \showarticletitle{Evolutionary Improvement of Assertion Oracles}. In \bibinfo{booktitle}{\emph{Joint European Software Engineering Conference and Symposium on the Foundations of Software Engineering}}. \bibinfo{pages}{1178–1189}.
\newblock


\bibitem[Terragni et~al\mbox{.}(2025)]%
        {terragni2025future}
\bibfield{author}{\bibinfo{person}{Valerio Terragni}, \bibinfo{person}{Annie Vella}, \bibinfo{person}{Partha Roop}, {and} \bibinfo{person}{Kelly Blincoe}.} \bibinfo{year}{2025}\natexlab{}.
\newblock \showarticletitle{The Future of AI-Driven Software Engineering}.
\newblock \bibinfo{journal}{\emph{ACM Trans. Softw. Eng. Methodol.}} \bibinfo{volume}{34}, \bibinfo{number}{5} (\bibinfo{date}{Jan.} \bibinfo{year}{2025}).
\newblock
\urldef\tempurl%
\url{https://doi.org/10.1145/3715003}
\showDOI{\tempurl}


\bibitem[TheAlgorithms(2025)]%
        {AESEncryptionTestGitHub}
\bibfield{author}{\bibinfo{person}{TheAlgorithms}.} \bibinfo{year}{2025}\natexlab{}.
\newblock \bibinfo{booktitle}{\emph{AESEncryptionTest}}.
\newblock
\urldef\tempurl%
\url{https://github.com/TheAlgorithms/Java/blob/master/src/test/java/com/thealgorithms/ciphers/AESEncryptionTest.java}
\showURL{%
\tempurl}


\bibitem[\tool(2025)]%
        {tool}
\bibfield{author}{\bibinfo{person}{\tool}.} \bibinfo{year}{2025}\natexlab{}.
\newblock \bibinfo{booktitle}{\emph{\tool website}}.
\newblock
\urldef\tempurl%
\url{https://mr-coupler.github.io/}
\showURL{%
Retrieved September 2, 2025 from \tempurl}


\bibitem[\tool(2026)]%
        {toolZenodo}
\bibfield{author}{\bibinfo{person}{\tool}.} \bibinfo{year}{2026}\natexlab{}.
\newblock \bibinfo{booktitle}{\emph{\tool on Zenodo}}.
\newblock
\urldef\tempurl%
\url{https://doi.org/10.5281/zenodo.19438045}
\showURL{%
Retrieved April 2, 2026 from \tempurl}


\bibitem[Tsigkanos et~al\mbox{.}(2023)]%
        {DBLP:conf/iccS/TsigkanosRMK23}
\bibfield{author}{\bibinfo{person}{Christos Tsigkanos}, \bibinfo{person}{Pooja Rani}, \bibinfo{person}{Sebastian M{\"{u}}ller}, {and} \bibinfo{person}{Timo Kehrer}.} \bibinfo{year}{2023}\natexlab{}.
\newblock \showarticletitle{Variable Discovery with Large Language Models for Metamorphic Testing of Scientific Software}. In \bibinfo{booktitle}{\emph{Computational Science - {ICCS} 2023 - 23rd International Conference, Prague, Czech Republic, July 3-5, 2023, Proceedings, Part {I}}} \emph{(\bibinfo{series}{Lecture Notes in Computer Science}, Vol.~\bibinfo{volume}{14073})}. \bibinfo{publisher}{Springer}, \bibinfo{pages}{321--335}.
\newblock


\bibitem[vmgama(2025)]%
        {dubboGitHub}
\bibfield{author}{\bibinfo{person}{vmgama}.} \bibinfo{year}{2025}\natexlab{}.
\newblock \bibinfo{booktitle}{\emph{dubbo}}.
\newblock
\urldef\tempurl%
\url{https://github.com/vmgama/dubbo/blob/master/dubbo-common/src/main/java/org/apache/dubbo/common/io/Bytes.java}
\showURL{%
\tempurl}


\bibitem[Wang et~al\mbox{.}(2020)]%
        {DBLP:conf/icsm/Wang0HSX0WL20}
\bibfield{author}{\bibinfo{person}{Ying Wang}, \bibinfo{person}{Bihuan Chen}, \bibinfo{person}{Kaifeng Huang}, \bibinfo{person}{Bowen Shi}, \bibinfo{person}{Congying Xu}, \bibinfo{person}{Xin Peng}, \bibinfo{person}{Yijian Wu}, {and} \bibinfo{person}{Yang Liu}.} \bibinfo{year}{2020}\natexlab{}.
\newblock \showarticletitle{An Empirical Study of Usages, Updates and Risks of Third-Party Libraries in Java Projects}. In \bibinfo{booktitle}{\emph{International Conference on Software Maintenance and Evolution}}. \bibinfo{publisher}{{IEEE}}, \bibinfo{pages}{35--45}.
\newblock


\bibitem[Wooldridge(2025)]%
        {sparseBitSetIssue13}
\bibfield{author}{\bibinfo{person}{Brett Wooldridge}.} \bibinfo{year}{2025}\natexlab{}.
\newblock \bibinfo{booktitle}{\emph{SparseBitSet Issue \#13}}.
\newblock
\urldef\tempurl%
\url{https://github.com/brettwooldridge/SparseBitSet/issues/13}
\showURL{%
\tempurl}


\bibitem[Xia et~al\mbox{.}(2024)]%
        {ICSE24_Fuzz4All}
\bibfield{author}{\bibinfo{person}{Chunqiu~Steven Xia}, \bibinfo{person}{Matteo Paltenghi}, \bibinfo{person}{Jia~Le Tian}, \bibinfo{person}{Michael Pradel}, {and} \bibinfo{person}{Lingming Zhang}.} \bibinfo{year}{2024}\natexlab{}.
\newblock \showarticletitle{Fuzz4All: Universal Fuzzing with Large Language Models}. In \bibinfo{booktitle}{\emph{International Conference on Software Engineering}}. \bibinfo{publisher}{{ACM}}, \bibinfo{pages}{126:1--126:13}.
\newblock


\bibitem[Xie et~al\mbox{.}(2023)]%
        {csqMTqa}
\bibfield{author}{\bibinfo{person}{Xiaoyuan Xie}, \bibinfo{person}{Shuo Jin}, {and} \bibinfo{person}{Songqiang Chen}.} \bibinfo{year}{2023}\natexlab{}.
\newblock \showarticletitle{qaAskeR\({}^{\mbox{+}}\): a novel testing method for question answering software via asking recursive questions}.
\newblock \bibinfo{journal}{\emph{Automated Software Engineering}} \bibinfo{volume}{30}, \bibinfo{number}{1} (\bibinfo{year}{2023}), \bibinfo{pages}{14}.
\newblock


\bibitem[Xie et~al\mbox{.}(2024)]%
        {csqMTtranslator}
\bibfield{author}{\bibinfo{person}{Xiaoyuan Xie}, \bibinfo{person}{Shuo Jin}, \bibinfo{person}{Songqiang Chen}, {and} \bibinfo{person}{Shing-Chi Cheung}.} \bibinfo{year}{2024}\natexlab{}.
\newblock \showarticletitle{Word Closure-Based Metamorphic Testing for Machine Translation}.
\newblock \bibinfo{journal}{\emph{ACM Transactions on Software Engineering and Methodology}} (\bibinfo{date}{jul} \bibinfo{year}{2024}).
\newblock


\bibitem[Xu et~al\mbox{.}(2024a)]%
        {mradopt}
\bibfield{author}{\bibinfo{person}{Congying Xu}, \bibinfo{person}{Songqiang Chen}, \bibinfo{person}{Jiarong Wu}, \bibinfo{person}{Shing{-}Chi Cheung}, \bibinfo{person}{Valerio Terragni}, \bibinfo{person}{Hengcheng Zhu}, {and} \bibinfo{person}{Jialun Cao}.} \bibinfo{year}{2024}\natexlab{a}.
\newblock \showarticletitle{MR-Adopt: Automatic Deduction of Input Transformation Function for Metamorphic Testing}. In \bibinfo{booktitle}{\emph{Proceedings of the 39th {IEEE/ACM} International Conference on Automated Software Engineering, {ASE} 2024, Sacramento, CA, USA, October 27 - November 1, 2024}}, \bibfield{editor}{\bibinfo{person}{Vladimir Filkov}, \bibinfo{person}{Baishakhi Ray}, {and} \bibinfo{person}{Minghui Zhou}} (Eds.). \bibinfo{publisher}{{ACM}}, \bibinfo{pages}{557--569}.
\newblock
\urldef\tempurl%
\url{https://doi.org/10.1145/3691620.3696020}
\showDOI{\tempurl}


\bibitem[Xu et~al\mbox{.}(2024b)]%
        {mrscout}
\bibfield{author}{\bibinfo{person}{Congying Xu}, \bibinfo{person}{Valerio Terragni}, \bibinfo{person}{Hengcheng Zhu}, \bibinfo{person}{Jiarong Wu}, {and} \bibinfo{person}{Shing{-}Chi Cheung}.} \bibinfo{year}{2024}\natexlab{b}.
\newblock \showarticletitle{MR-Scout: Automated Synthesis of Metamorphic Relations from Existing Test Cases}.
\newblock \bibinfo{journal}{\emph{ACM Transactions on Software Engineering and Methodology}} \bibinfo{volume}{33}, \bibinfo{number}{6} (\bibinfo{year}{2024}), \bibinfo{pages}{150}.
\newblock


\bibitem[Yang et~al\mbox{.}(2024)]%
        {DBLP:journals/corr/abs-2404-14646}
\bibfield{author}{\bibinfo{person}{Zhen Yang}, \bibinfo{person}{Fang Liu}, \bibinfo{person}{Zhongxing Yu}, \bibinfo{person}{Jacky~Wai Keung}, \bibinfo{person}{Jia Li}, \bibinfo{person}{Shuo Liu}, \bibinfo{person}{Yifan Hong}, \bibinfo{person}{Xiaoxue Ma}, \bibinfo{person}{Zhi Jin}, {and} \bibinfo{person}{Ge Li}.} \bibinfo{year}{2024}\natexlab{}.
\newblock \showarticletitle{Exploring and Unleashing the Power of Large Language Models in Automated Code Translation}.
\newblock \bibinfo{journal}{\emph{CoRR}}  \bibinfo{volume}{abs/2404.14646} (\bibinfo{year}{2024}).
\newblock
\showeprint[arXiv]{2404.14646}


\bibitem[Yuan et~al\mbox{.}(2021)]%
        {shuaiMTqa}
\bibfield{author}{\bibinfo{person}{Yuanyuan Yuan}, \bibinfo{person}{Shuai Wang}, \bibinfo{person}{Mingyue Jiang}, {and} \bibinfo{person}{Tsong~Yueh Chen}.} \bibinfo{year}{2021}\natexlab{}.
\newblock \showarticletitle{Perception Matters: Detecting Perception Failures of {VQA} Models Using Metamorphic Testing}. In \bibinfo{booktitle}{\emph{Conference on Computer Vision and Pattern Recognition}}. \bibinfo{publisher}{Computer Vision Foundation / {IEEE}}, \bibinfo{pages}{16908--16917}.
\newblock


\bibitem[Yuan et~al\mbox{.}(2024)]%
        {arxiv23_FDUGPT4TestGen}
\bibfield{author}{\bibinfo{person}{Zhiqiang Yuan}, \bibinfo{person}{Mingwei Liu}, \bibinfo{person}{Shiji Ding}, \bibinfo{person}{Kaixin Wang}, \bibinfo{person}{Yixuan Chen}, \bibinfo{person}{Xin Peng}, {and} \bibinfo{person}{Yiling Lou}.} \bibinfo{year}{2024}\natexlab{}.
\newblock \showarticletitle{Evaluating and Improving ChatGPT for Unit Test Generation}.
\newblock \bibinfo{journal}{\emph{Proc. {ACM} Softw. Eng.}} \bibinfo{volume}{1}, \bibinfo{number}{{FSE}} (\bibinfo{year}{2024}), \bibinfo{pages}{1703--1726}.
\newblock
\urldef\tempurl%
\url{https://doi.org/10.1145/3660783}
\showDOI{\tempurl}


\bibitem[Zhang et~al\mbox{.}(2019)]%
        {zhang2019automatic}
\bibfield{author}{\bibinfo{person}{Bo Zhang}, \bibinfo{person}{Hongyu Zhang}, \bibinfo{person}{Junjie Chen}, \bibinfo{person}{Dan Hao}, {and} \bibinfo{person}{Pablo Moscato}.} \bibinfo{year}{2019}\natexlab{}.
\newblock \showarticletitle{Automatic Discovery and Cleansing of Numerical Metamorphic Relations}. In \bibinfo{booktitle}{\emph{{IEEE} International Conference on Software Maintenance and Evolution}}. \bibinfo{publisher}{{IEEE}}, \bibinfo{pages}{235--245}.
\newblock


\bibitem[Zhang et~al\mbox{.}(2014)]%
        {zhang2014search}
\bibfield{author}{\bibinfo{person}{Jie Zhang}, \bibinfo{person}{Junjie Chen}, \bibinfo{person}{Dan Hao}, \bibinfo{person}{Yingfei Xiong}, \bibinfo{person}{Bing Xie}, \bibinfo{person}{Lu Zhang}, {and} \bibinfo{person}{Hong Mei}.} \bibinfo{year}{2014}\natexlab{}.
\newblock \showarticletitle{Search-based inference of polynomial metamorphic relations}. In \bibinfo{booktitle}{\emph{{ACM/IEEE} International Conference on Automated Software Engineering}}. \bibinfo{publisher}{{ACM}}, \bibinfo{pages}{701--712}.
\newblock


\bibitem[Zhang et~al\mbox{.}(2025b)]%
        {DBLP:conf/saner/ZhangSLD25}
\bibfield{author}{\bibinfo{person}{Jiaming Zhang}, \bibinfo{person}{Chang{-}Ai Sun}, \bibinfo{person}{Huai Liu}, {and} \bibinfo{person}{Sijin Dong}.} \bibinfo{year}{2025}\natexlab{b}.
\newblock \showarticletitle{Can Large Language Models Discover Metamorphic Relations? {A} Large-Scale Empirical Study}. In \bibinfo{booktitle}{\emph{{IEEE} International Conference on Software Analysis, Evolution and Reengineering, {SANER} 2025, Montreal, QC, Canada, March 4-7, 2025}}. \bibinfo{publisher}{{IEEE}}, \bibinfo{pages}{24--35}.
\newblock
\urldef\tempurl%
\url{https://doi.org/10.1109/SANER64311.2025.00011}
\showDOI{\tempurl}


\bibitem[Zhang et~al\mbox{.}(2025a)]%
        {DBLP:journals/infsof/ZhangCPTYZ25}
\bibfield{author}{\bibinfo{person}{Yifan Zhang}, \bibinfo{person}{Tsong~Yueh Chen}, \bibinfo{person}{Matthew Pike}, \bibinfo{person}{Dave Towey}, \bibinfo{person}{Zhihao Ying}, {and} \bibinfo{person}{Zhi~Quan Zhou}.} \bibinfo{year}{2025}\natexlab{a}.
\newblock \showarticletitle{Enhancing autonomous driving simulations: {A} hybrid metamorphic testing framework with metamorphic relations generated by {GPT}}.
\newblock \bibinfo{journal}{\emph{Inf. Softw. Technol.}}  \bibinfo{volume}{187} (\bibinfo{year}{2025}), \bibinfo{pages}{107828}.
\newblock
\urldef\tempurl%
\url{https://doi.org/10.1016/J.INFSOF.2025.107828}
\showDOI{\tempurl}


\bibitem[Zhang et~al\mbox{.}(2023)]%
        {DBLP:conf/compsac/ZhangTP23}
\bibfield{author}{\bibinfo{person}{Yifan Zhang}, \bibinfo{person}{Dave Towey}, {and} \bibinfo{person}{Matthew Pike}.} \bibinfo{year}{2023}\natexlab{}.
\newblock \showarticletitle{Automated Metamorphic-Relation Generation with ChatGPT: An Experience Report}. In \bibinfo{booktitle}{\emph{47th {IEEE} Annual Computers, Software, and Applications Conference, {COMPSAC} 2023, Torino, Italy, June 26-30, 2023}}. \bibinfo{publisher}{{IEEE}}, \bibinfo{pages}{1780--1785}.
\newblock
\urldef\tempurl%
\url{https://doi.org/10.1109/COMPSAC57700.2023.00275}
\showDOI{\tempurl}


\bibitem[Zhang et~al\mbox{.}(2025c)]%
        {zhang2025integrating}
\bibfield{author}{\bibinfo{person}{Yifan Zhang}, \bibinfo{person}{Dave Towey}, \bibinfo{person}{Matthew Pike}, \bibinfo{person}{Quang{-}Hung Luu}, \bibinfo{person}{Huai Liu}, {and} \bibinfo{person}{Tsong~Yueh Chen}.} \bibinfo{year}{2025}\natexlab{c}.
\newblock \showarticletitle{Integrating Artificial Intelligence with Human Expertise: {A}n In-depth Analysis of ChatGPT's Capabilities in Generating Metamorphic Relations}.
\newblock \bibinfo{journal}{\emph{CoRR}}  \bibinfo{volume}{abs/2503.22141} (\bibinfo{year}{2025}).
\newblock
\showeprint[arXiv]{2503.22141}
\urldef\tempurl%
\url{https://arxiv.org/abs/2503.22141}
\showURL{%
\tempurl}


\bibitem[Zhang et~al\mbox{.}(2025d)]%
        {llmhallucination}
\bibfield{author}{\bibinfo{person}{Ziyao Zhang}, \bibinfo{person}{Chong Wang}, \bibinfo{person}{Yanlin Wang}, \bibinfo{person}{Ensheng Shi}, \bibinfo{person}{Yuchi Ma}, \bibinfo{person}{Wanjun Zhong}, \bibinfo{person}{Jiachi Chen}, \bibinfo{person}{Mingzhi Mao}, {and} \bibinfo{person}{Zibin Zheng}.} \bibinfo{year}{2025}\natexlab{d}.
\newblock \showarticletitle{{LLM} Hallucinations in Practical Code Generation: Phenomena, Mechanism, and Mitigation}.
\newblock \bibinfo{journal}{\emph{Proc. {ACM} Softw. Eng.}} \bibinfo{volume}{2}, \bibinfo{number}{{ISSTA}} (\bibinfo{year}{2025}), \bibinfo{pages}{481--503}.
\newblock
\urldef\tempurl%
\url{https://doi.org/10.1145/3728894}
\showDOI{\tempurl}


\bibitem[Zhou et~al\mbox{.}(2020)]%
        {zhou2018metamorphic}
\bibfield{author}{\bibinfo{person}{Zhi~Quan Zhou}, \bibinfo{person}{Liqun Sun}, \bibinfo{person}{Tsong~Yueh Chen}, {and} \bibinfo{person}{Dave Towey}.} \bibinfo{year}{2020}\natexlab{}.
\newblock \showarticletitle{Metamorphic Relations for Enhancing System Understanding and Use}.
\newblock \bibinfo{journal}{\emph{IEEE Transactions on Software Engineering}} \bibinfo{volume}{46}, \bibinfo{number}{10} (\bibinfo{year}{2020}), \bibinfo{pages}{1120--1154}.
\newblock


\bibitem[Zingg(2025)]%
        {zinggIssue60}
\bibfield{author}{\bibinfo{person}{Zingg}.} \bibinfo{year}{2025}\natexlab{}.
\newblock \bibinfo{booktitle}{\emph{Zingg Issue \#60}}.
\newblock
\urldef\tempurl%
\url{https://github.com/zinggAI/zingg/issues/60}
\showURL{%
\tempurl}


\end{thebibliography}

\end{document}